\documentclass[letterpaper]{article}

\usepackage[T1]{fontenc}

\usepackage{geometry}
\usepackage{setspace}

\usepackage[
  backend=biber,
  style=chem-acs,
  articletitle,
  date=year,
  url=false
]{biblatex}
\DeclareBibliographyDriver{online}{%
  \usebibmacro{bibindex}%
  \usebibmacro{begentry}%
  \usebibmacro{author}%
  \setunit{\labelnamepunct}\newblock
  \usebibmacro{title}%
  \newunit\newblock
  \printdate%
  \newunit\newblock
  \usebibmacro{url+urldate}%
  \usebibmacro{finentry}%
}
\DeclareCiteCommand{\citeplain}
  {\usebibmacro{prenote}}
  {\usebibmacro{citeindex}%
   \printtext[bibhyperref]{\printfield{labelnumber}}}
  {\multicitedelim}
  {\usebibmacro{postnote}}

\usepackage{graphicx}
\usepackage{float}
\usepackage{multirow}%
\usepackage{amsmath,amssymb,amsfonts}%
\usepackage{amsthm}%
\usepackage{mathrsfs}%
\usepackage[title]{appendix}%
\usepackage{xcolor}%
\usepackage{textcomp}%
\usepackage{manyfoot}%
\usepackage{booktabs}%
\usepackage{algorithm}%
\usepackage{algorithmicx}%
\usepackage{algpseudocode}%
\usepackage{listings}%
\usepackage{braket}
\usepackage{siunitx}
\usepackage{hyperref}

\newfloat{scheme}{htbp}{los}
\floatname{scheme}{Scheme}
\floatname{chart}{Chart}
\newfloat{graph}{htbp}{loh}

\usepackage{pdfpages} 
\usepackage{pgffor} 

\makeatletter
\AtBeginDocument{\let\LS@rot\@undefined}
\makeatother

\def\supplementfilename{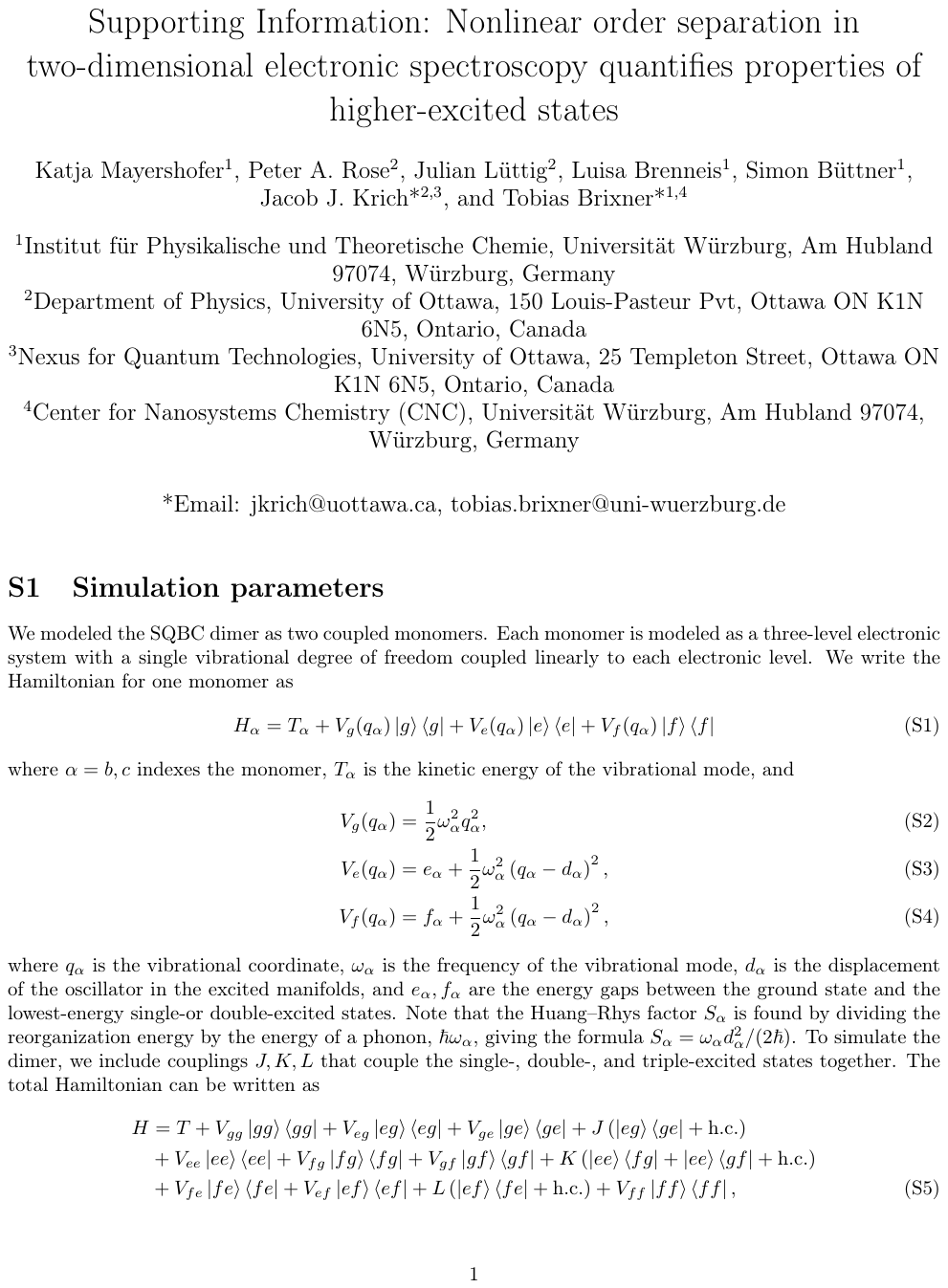}

\pdfximage{\supplementfilename}
\def\numbersupplementpages{\the\pdflastximagepages}

\newif\ifarXiv
\arXivtrue 

\usepackage{authblk}
\author[1]{Katja Mayershofer}
\author[2]{Peter A. Rose}
\author[2]{Julian Lüttig}
\author[1]{Luisa Brenneis}
\author[1]{Simon Büttner}
\author[2,3]{Jacob J. Krich*}
\author[1,4]{Tobias Brixner*}

\affil[1]{Institut für Physikalische und Theoretische Chemie, Universität Würzburg, Am Hubland 97074, Würzburg, Germany}
\affil[2]{Department of Physics, University of Ottawa, 150 Louis-Pasteur Pvt, Ottawa ON K1N 6N5, Ontario, Canada}
\affil[3]{Nexus for Quantum Technologies, University of Ottawa, 25 Templeton Street, Ottawa ON K1N 6N5, Ontario, Canada}
\affil[4]{Center for Nanosystems Chemistry (CNC), Universität Würzburg, Am Hubland 97074, Würzburg, Germany}

\title{Nonlinear order separation in two-dimensional electronic spectroscopy quantifies properties of higher-excited states}
\date{*Email: jkrich@uottawa.ca, tobias.brixner@uni-wuerzburg.de}

\begin{document}

\maketitle

\newpage
\begin{abstract}
  Two-dimensional (2D) spectroscopy combines high temporal and spectral resolution, allowing the observation of ultrafast energy transfer and the separation of homogeneous and inhomogeneous broadening. Typically, 2D spectroscopy is dominated by the lowest-order nonlinear signal for a given phase-matching configuration while signals of higher order are present but difficult to access separately. Recently, we introduced a technique to separate nonlinear orders in 2D spectroscopy by systematically varying the intensity of the pump pulses and appropriate post-processing. Here, we unravel the full potential of higher-order 2D spectroscopy by separating multiple nonlinear orders at different multi-quantum positions. As an example, we investigate a squaraine dimer. Using a theoretical model, we find excellent qualitative and quantitative agreement throughout all nonlinear orders and multi-quantum positions. Our simulations demonstrate the sensitivity and information content hidden in the higher-order spectra such as transition dipole moments and energy levels even of highly excited states. Our results pave the way for establishing higher-order spectroscopy as a unique extension of multidimensional spectroscopy, providing access to highly excited states and their properties encoded in successive orders of nonlinearity.
\end{abstract}

\section*{TOC Graphic}

\begin{figure}[h]
\centering
\includegraphics[width=0.9\textwidth]{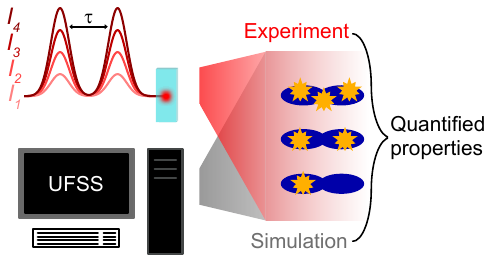}

\end{figure}

\newpage
\section{Introduction}

Two-dimensional (2D) spectroscopy is an established method that combines ultrafast temporal resolution and multidimensional spectral information \cite{hamm_concepts_2011-1, fuller_experimental_2015}. As a generalization of transient absorption (TA) with resolution along both excitation- and detection-frequency axes, 2D spectroscopy is frequently used to investigate dynamics of processes in complex quantum systems. Examples include, but are not limited to, energy transfer processes in photosynthetic systems \cite{brixner_two-dimensional_2005, dostal_situ_2016, silori_two-dimensional_2023, akhtar_ultrafast_2024,keil_ultrafast_2026}, coherences between states \cite{thyrhaug_identification_2018, policht_hidden_2022}, chemical reactions \cite{ruetzel_tracing_2013, ruetzel_multidimensional_2014}, charge-transfer state dynamics \cite{mandal_transient_2019}, singlet fission processes \cite{bakulin_real-time_2016}, and many more, as summarized in several review articles \cite{cho_coherent_2008, tiwari_multidimensional_2021, biswas_coherent_2022, gelzinis_two-dimensional_2019, fresch_two-dimensional_2023,maiuri_ultrafast_2020, fiebig_ultrafast_2023}. One of the key features of 2D spectroscopy is the separation of signal contributions such as ground-state bleach (GSB), stimulated emission (SE), and excited-state absorption (ESA) along the excitation and detection axes, allowing the dynamics and properties of individual excited states to be probed. 

A frequent problem in spectroscopic experiments is to find a balance between a sufficient signal-to-noise ratio (SNR) and overlapping higher-order contributions. The most commonly used TA and 2D experiments investigate third-order signals, in which the pump pulses interact twice with the system and the probe once \cite{hamm_concepts_2011-1, branczyk_crossing_2014}; an increase in pump intensity leads to better SNR, but also increases higher-order contributions originating from more than two interactions with the pump pulses. These higher-order signals overlap with the desired third-order signal \cite{muller_singlet_2010-1, volker_singletsinglet_2014, lee_ultrafast_2018}. Higher-order contributions contain new information that is not part of a common third-order 2D spectrum, such as exciton--exciton interaction \cite{rehhagen_exciton_2020, dahlberg_mapping_2017} and properties of higher-excited states \cite{maly_signatures_2020,rose_interpretations_2023,dostal_direct_2018}. However, because the lower-order and higher-order signals are mixed, such information is difficult to extract. Multi-quantum 2D spectroscopy experiments have signals whose lowest-order contributions result from more than two interactions with an excitation pulse, but contributions by even higher orders in nonlinearity can still contaminate the desired signal \cite{bruggemann_nonperturbative_2011,dostal_direct_2018, maly_wavelike_2020, heshmatpour_annihilation_2020}. The difficulty of separating higher-order signals and accounting for their contributions is a common challenge in spectroscopy, regardless of technique or sample.

We solved this longstanding problem in pump--probe spectroscopy such as TA by introducing the technique of intensity cycling, which systematically separates the nonlinear orders of response by using linear combinations of measurements at several excitation intensities \cite{maly_separating_2023, luttig_high-order_2023, luttig_higher-order_2023-1}. Recently, we generalized that approach and demonstrated that variation of intensity of excitation pulses allows nonlinear higher-order contributions to be extracted in 2D spectroscopy and related techniques \cite{krich_separating_2025}. The now separable nonlinear order responses contain spectral and dynamical information about both singly \cite{krich_separating_2025,rose_interpretations_2023} and multiply \cite{maly_separating_2023,rose_interpretations_2023} excited states. However, it is an open question what types of information one can extract from such higher-order spectra and to what extent quantitative information about quantum-mechanical properties can be determined. In the past, the presence of various types of artifact, such as the uncontrolled presence of higher orders, often precluded a quantitative analysis for complex systems.

In this work, we show how single-quantum and double-quantum 2D electronic spectra, resolved into the various orders of nonlinear response, can be analyzed quantitatively by comparing accurate experiments and simulations. Specifically, we obtain high-order 2D spectra of a squaraine dimer and compare them to simulations that include the effects of the experimental pulse shape. The removal of artifacts arising from higher orders allows us to obtain quantitative insight. From the linear absorption and third-order response, we retrieve energy levels, relative dipole couplings of the singly excited states, and biexciton binding energies. From the higher orders, we determine the couplings of the single-exciton states to the doubly excited states. We thus produce a well-constrained model of this exemplary dimer. This work demonstrates the route to quantitative analysis of 2D electronic spectra.

\section{Results and discussion}

We extract nonlinear orders through intensity variation across different $n$-quantum ($n$Q) positions in the 2D spectra. With noncollinear pulses, the $n$Q signals are detected in the $-n\vec{k}_1+n\vec{k}_2+\vec{k}_3$ direction. Here, we use a multidimensional experiment in the pump--probe geometry with $\vec{k}_1=\vec{k}_2$ and thus all $n$Q signals are emitted in the same phase-matching direction. However, the signals can be separated along the $\omega_\tau$ axis, where $\tau$ is the delay between pump pulses; the $n$Q spectra appear centered at $\omega_\tau = n\omega_0$, where $\omega_0$ is the central frequency of the excitation pulses \cite{luttig_higher-order_2023-1}. The lowest-order contributions to the $n$Q spectra are of $2n^\text{th}$ order in the pump pulse amplitudes. To separate nonlinear orders at all $n$Q positions, we first determine the optimal set of pump intensities, following the procedure described by Krich et al. \cite{krich_separating_2025}. The 2D measurements are then taken for the optimized intensity steps and the resulting 2D data are used to isolate the signals of different orders of nonlinear response.

\begin{figure}
\centering
\includegraphics[width=0.9\textwidth]{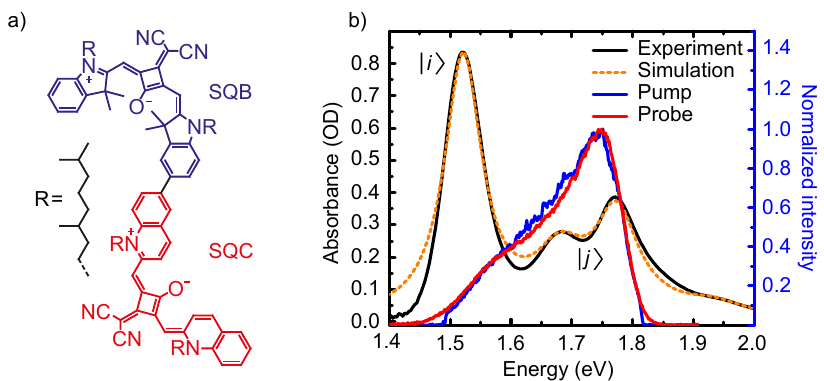}
\caption{Characteristics of the investigated squaraine dimer dSQBC. a)~Molecular structure of dSQBC with the SQB part in blue and the SQC part in red. b)~Experimental (black line) and simulated (orange, dashed line) absorption spectra in toluene; pump (blue) and probe (red) laser spectra.}
\label{fig1_Absorp_Molec}
\end{figure}

To demonstrate the principle experimentally, we dissolved a squaraine dimer, dSQBC, which is composed of an SQB and an SQC monomer (molecular structure shown in Figure~\ref{fig1_Absorp_Molec}a), in toluene \cite{schreck_synthesis_2018-1}. Figure~\ref{fig1_Absorp_Molec}b shows the absorption spectrum in black, as well as the pump and probe spectra in blue and red, respectively. For comparison, Figure~\ref{fig1_Absorp_Molec}b also shows the simulated linear absorption, described below, as a dashed orange line. The experimental linear spectrum consists of a large peak at \qty{1.52}{eV} and two overlapping peaks at \qty{1.68}{eV} and \qty{1.78}{eV}. The peak at \qty{1.52}{eV} corresponds to a transition to an excitonic state in the vibrational ground state, which we will call $\ket{i}$, while the two overlapping peaks indicate a vibronic mixture. This second exciton species will be referred to as $\ket{j}$. The squaraine dimer dSQBC shows slight changes in the relative peak strength for different sample concentrations, as seen in Figure~S12. For the simulations, we do not take this behavior into account, as the effects are minimal and larger aggregation effects can be ruled out, as there are no energetically shifted peaks in the differently concentrated spectra.  The pump and probe spectra have identical shapes and cover all three peaks of the absorption spectrum. Due to the asymmetric shape of our laser spectra, the excitation intensity is enhanced at the blue edge of the absorption spectrum.

For the intensity-varied measurements we can define the electric field at time $t$ and position $\vec{r}$ produced by the three pulses in the 2D measurement as 
\begin{equation}
    \vec{E}\left(t,\vec{r} \right)=\sum_{l=1}^{3}\lambda_l\vec{e}_l\epsilon_l\left(t-\vec{k}_l\cdot\vec{r}/\omega_l\right)e^{-i\left(\omega_l\left(t-t_l\right)-\vec{k}_l\cdot\vec{r}\right)}+\text{c.c.},
    \label{eqEfield}
\end{equation}
where $\epsilon_l$ describes the complex pulse envelope, including any chirp, $\omega_l$ the central frequency,  $t_l$ the arrival time, and $\vec{e}_l$ the polarization of the pulse number $l$. The scaling factor $\lambda_l$ allows varying the amplitude while leaving all other pulse parameters unchanged. For our intensity-varied measurements, $\lambda_3$ is always $1$ for the probe pulse and $\lambda_1=\lambda_2\equiv\lambda$ for the pump pulses. We then define the dimensionless intensity $I=\lambda^2$ that describes the simultaneous scaling of the amplitudes of the excitation pulse pair. The 2D spectroscopy signal $S(\tau,T,t,I)$ can then be written as 
\begin{equation}
 S\left(\tau,T,t,I\right)=\sum_{m=1}^\infty S^{(2m+1)}\left(\tau,T,t\right)I^m,\label{eqSn(I)}
\end{equation}
where $\tau$ and $T$ are the interpulse delays \cite{krich_separating_2025}. Here, $S\left(\tau,T,t,I\right)$ is the intensity-dependent signal and $S^{(2m+1)}\left(\tau,T,t\right)$ is the signal that is of $2m^\text{th}$ order in the excitation-pulse amplitudes. Since $\lambda$ scales the excitation-pulse amplitudes and $I=\lambda^2$, the $S^{(2m+1)}$ signals are scaled by $I^m$ because the excitation-pulse amplitudes interact $2m$ times. As shown elsewhere \cite{krich_separating_2025}, we can extract the first $K+1$ orders in Eq.~\eqref{eqSn(I)} by measuring $S\left(\tau,T,t,I\right)$ with $K$ different values of $I$ and truncating the sum at an upper limit of $m=K$ instead of $m=\infty$. This truncation produces a systematic error, originating from the terms $S^{(2m+1)}$ for $m>K$. This systematic error can be estimated, and therefore the total error consisting of the random error and the systematic error can be minimized by an appropriate choice of $K$ excitation intensities $\{I_{k}\}$. In this work, we use the same experimental procedure as in the proof-of-principle study on another sample \cite{krich_separating_2025} but then focus on how quantitative information about the system can be extracted when accurate experiments and simulations are combined.

\begin{figure}[h]
\centering
\includegraphics[width=0.9\textwidth]{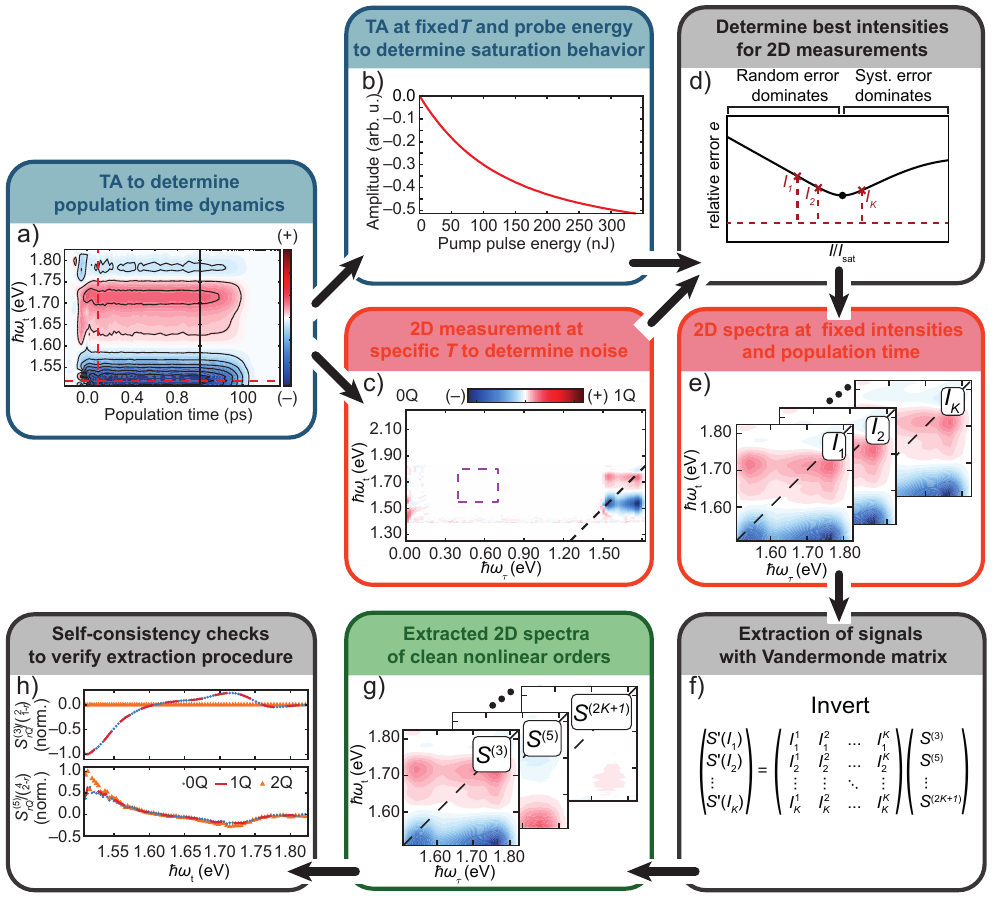}
\caption{Scheme showing the experimental and data processing steps to separate signals of different orders of response. TA measurements are shown as blue boxes, 2D measurements as red boxes, data analysis steps as gray boxes, and the extracted 2D data as a green box.}\label{fig2_MethodScheme}
\end{figure}

Figure~\ref{fig2_MethodScheme} shows an overview of the steps necessary to extract the nonlinear signals. In the first step, we analyzed the population-time dynamics in a TA experiment to determine what population time to use in our intensity-varied 2D measurement. In general, our method also works for experiments with multiple population times. However, to keep the measurement time short and decrease any potential influence from laser instabilities during the measurement, we decided to only analyze 2D data from a single population time. We used a weak probe beam (with an intensity of a fourth of the lowest pump intensity), so the probe only interacts once. Figure \ref{fig2_MethodScheme}a shows the TA measurement of dSQBC. From this measurement, we chose a population time of \qty{100}{fs}, marked by the vertical red dashed line in Figure~\ref{fig2_MethodScheme}a, because the TA signal exhibits a maximum signal intensity at around \qty{100}{fs} and pulse overlap effects are no longer present at this population time in the 2D measurements. Furthermore, time-dependent 2D measurements also showed that the 2Q signal displays its maximum signal intensity at \qty{100}{fs}. The early-time dynamics of the TA signal are shown in more detail in Supporting Information Figure~S1.

In the next steps, the ideal intensities for the extraction of response orders in the 2D spectra were determined. First, an estimate of the systematic error in the 2D spectra was obtained by studying the saturation behavior of the sample in an intensity-dependent TA measurement \cite{krich_separating_2025}, based on the fact that the TA spectrum is the excitation-frequency integral of the 2D spectrum according to the projection-slice theorem \cite{jonas_two-dimensional_2003}. A TA experiment (Figure~\ref{fig2_MethodScheme}b) was measured with pump intensities ranging from \qty{0.4}{nJ} to \qty{339}{nJ}, with the spatial and temporal envelope of the pump held constant, so that the intensity and energy of the pulse were proportional. The intensity-dependent TA signal $S^{\text{TA}}$ was fitted using the exponential saturation form
\begin{equation}
 S^{\text{TA}}\left (T,\omega_{t} ,I \right )=-S_{\text{max}}\left ( T,\omega _{t} \right )\left ( 1-e^{-I/I_{\text{sat}}\left ( T,\omega _{t} \right )} \right ),\label{eqSatTA}
\end{equation}
where $S^{\text{TA}}\left (T,\omega_{t} ,I \right )$ is the TA signal at population time $T$, frequency $\omega_{t}$, and pump intensity $I$, and $S_{\text{max}}$ and $I_\text{sat}$ are sample- and pulse-dependent characteristics of the saturation form. From this fit, shown in more detail in Supporting Information Section~S3, we retrieved the values of $I_{\text{sat}}=\qty{124}{nJ}$ and $S_{\text{max}}=0.53$ at $T=\qty{100}{fs}$ and $\omega_{t}=\qty{1.51}{eV}$, which we used to estimate the systematic error in 2D measurements.

We also must quantify the noise level in the 2D measurements to find the optimal excitation intensities. To determine the noise level, a 2D measurement was taken at the population time $T=\qty{100}{fs}$, and the noise level was determined far from any expected signal, in the region of interest marked by the purple square on the 2D map in Figure~\ref{fig2_MethodScheme}c. For our measurement a value of $5\times10^{-7}$ was calculated for the standard deviation of the noise following the procedure outlined in the Supporting Information of our previous work \cite{krich_separating_2025}.

At this point, both main sources of error were known: systematic error due to saturation behavior and random error due to noise in the 2D measurement. The combination of these two types of error was then used to calculate the optimal pump intensities to minimize the relative error in the extracted orders (Figure~\ref{fig2_MethodScheme}d). In our case, the experimental setup limited the maximum possible intensity of the excitation pulse to \qty{15}{nJ}, which led to an ideal number of intensity steps as described in \citeplain[Ref.][]{krich_separating_2025}. From the error analysis, four ideal intensities were determined, corresponding to pump pulse energies of \qty{1.74}{nJ}, \qty{7.35}{nJ}, \qty{12.9}{nJ}, and \qty{15}{nJ}. These are optimal for minimizing the error in the first three nonlinear orders $S^{(3)}$, $S^{(5)}$, and $S^{(7)}$, i.e., the third- to seventh-order responses, at both 1Q and 2Q positions. The code for finding the optimal intensities is freely available on GitHub \cite{jkrich_jkrichintensity_optimization_2026}.

The 2D measurements were then performed at the four ideal intensities calculated and the chosen population time (Figure~\ref{fig2_MethodScheme}e). The complete 2D maps are shown in Supporting Information Figure~S4. The nonlinear orders were then extracted at each $\omega_\tau$ and $\omega_t$ by inversion of the equation
 \begin{equation}
 \begingroup
\renewcommand*{\arraystretch}{1.25}
 \begin{pmatrix}
S(I_{1})\\
S(I_{2}) \\
S(I_{3}) \\
S(I_{4})
\end{pmatrix}=\begin{pmatrix}
I_{1} & I_{1}^2 & I_{1}^3 &I_{1}^4\\
I_{2} & I_{2}^2 & I_{2}^3 &I_{2}^4\\
I_{3} & I_{3}^2 & I_{3}^3 &I_{3}^4\\
I_{4} & I_{4}^2 & I_{4}^3 &I_{4}^4\\
\end{pmatrix}\begin{pmatrix}
S^{(3)} \\
S^{(5)} \\
S^{(7)} \\
S^{(9)} \\
\end{pmatrix},\label{eqVdmMatrix}
\endgroup
\end{equation}
which is the truncated form of Eq.~\eqref{eqSn(I)} with $K=4$. The 2D spectra of nonlinear orders up to $S^{(9)}$ were extracted and are shown in Figure~\ref{fig2_MethodScheme}g.

In the final step of the extraction and measurement procedure, self-consistency checks were performed comparing the $n$Q signals located at different positions in the 2D spectrum \cite{krich_separating_2025}; for our experiment we compared the signals at the 0Q, 1Q, and 2Q positions (Figure~\ref{fig2_MethodScheme}h). The $S^{(3)}$ signals at the 0Q and 1Q positions show excellent agreement, and the $S^{(3)}$ 2Q signal shows no signal amplitude as expected, since the lowest nonlinear signal contributing at the 2Q position is $S^{(5)}$. For the $S^{(5)}$ signal, the integrated and scaled signals at all three $n$Q positions show good agreement, only showing some minor disagreement at the edge of the excitation spectrum. These self-consistency checks are shown in more detail in Supporting Information Section~S5; they give a good indication that the extracted orders $S^{(3)}$ and $S^{(5)}$ are of high quality, with minimal contamination error.

\begin{figure}[h]
\centering
\includegraphics[width=1.0\textwidth]{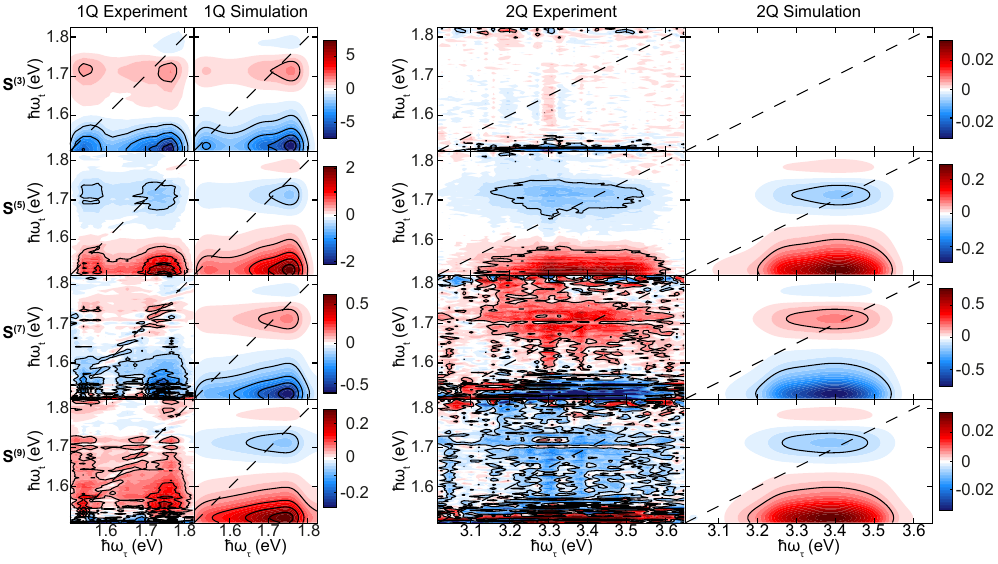}
\caption{Separated experimental and simulated 2D spectra of different nonlinear orders and quantum positions at the population time $T=\qty{100}{fs}$. The two left columns show the experiment and simulation of the 1Q maps, while the two columns on the right the experimental data and simulations for the 2Q signal. Each row represents a signal of a specific nonlinear order from  $S^{(3)}$ (top) to $S^{(9)}$ (bottom). The dashed lines indicate the diagonals at $\hbar\omega_{\tau}=\hbar\omega_{t}$ (1Q) and $\hbar\omega_{\tau}=2\hbar\omega_{t}$ (2Q).}\label{fig3_Exp_vs_Sim}
\end{figure}

We can now analyze the isolated spectral orders. As expected, the $S^{(3)}$ spectrum at the 1Q position, shown in Figure~\ref{fig3_Exp_vs_Sim} (top left), displays the same three peaks along the excitation frequency axis as the linear absorption spectrum, but weighted by the pump intensity, which causes the highest-energy peak to be the strongest. The two higher-energy peaks visible in the linear absorption spectrum mostly merge in the nonlinear spectrum; we refer to the peaks between $\hbar\omega_\tau=\qty{1.65}{eV}$ and \qty{1.8}{eV} as the $\ket{j}$ peak and the lowest-energy peak at $\hbar\omega_\tau=\qty{1.55}{eV}\equiv\hbar\omega_i$ as the $\ket{i}$ peak, corresponding to excitation of the lowest-energy exciton. Along the detection frequency axis, $S^{(3)}$ shows two distinct bands, a negative band between $\hbar\omega_{t}=\qty{1.52}{eV}$ and \qty{1.60}{eV} and a positive band between $\hbar\omega_{t}=\qty{1.65}{eV}$ and \qty{1.75}{eV}. The negative band indicates that it is dominated by GSB and SE from $\ket{i}$, while the higher-energy band is positive and dominated by ESA from $\ket{i}$ to the biexciton, which we label $\ket{ij}$. The ESA from $\ket{i}$ to $\ket{ij}$ is at nearly the same $\hbar\omega_t$ spectral position as the GSB of $\ket{j}$. 

The $S^{(3)}$ spectrum shows that excited populations have mostly relaxed to $\ket{i}$ before the probe arrives. The upper diagonal peak, corresponding to absorption and re-emission at $\hbar\omega_{\tau}=\hbar\omega_{t}=\qty{1.76}{eV}$, is only barely visible, indicating that the higher-excited exciton has mostly decayed and no longer displays SE from that location. Similarly, the mid-diagonal peak at $\hbar\omega_{\tau}=\hbar\omega_{t}=\qty{1.67}{eV}$ is not visible at all. Cuts taken at different $\hbar\omega_\tau$ positions, depicted in Supporting Information Figure~S7, also show that the system has predominantly relaxed to the lowest single-excited state by \qty{100}{fs}; the normalized linecuts overlap, indicating that even though different states are excited by the pump pulse, as seen by the multiple peaks along the $\omega_\tau$ axis, the probe interacts only with one state, as the same energetic structure is visible for all cuts along the $\omega_t$ axis. The TA data shown in Supporting Information Section~S2 further support the conclusion that the system is mostly relaxed by $T=\qty{100}{fs}$, and the SE appearing only at $\hbar\omega_t=\hbar\omega_i$ is consistent with emission coming from the lowest-energy exciton. 

The ESA band near $\omega_t=\omega_j$ reveals the energy gap between $\ket{i}$ and $\ket{ij}$ and the strength of the transition dipole moment connecting $\ket{i}$ to $\ket{ij}$. The $S^{(3)}$ ESA band does not contain information about transitions to the doubly excited states from higher-energy singly excited states because they are not occupied. Information about the higher-lying singly excited states could be extracted from $S^{(3)}$ alone by considering smaller population times, which would involve detailed modeling of dynamics and might also require considering population times where pump and probe pulses overlap, which can be difficult. 

The higher-order 1Q spectra (Figure~\ref{fig3_Exp_vs_Sim}, left column, second to fourth row) have the same peak structure as $S^{(3)}$ but with signs alternating as the order increases, with the band at higher $\omega_t$  obscured by noise for $S^{(7)}$ and $S^{(9)}$. This peak structure is consistent with the multiply excited states mostly having relaxed to the singly excited states, so there are no significant novel signals arising from multiply excited states. The relaxation of multi-excitons is consistent with the known fast timescales of exciton--exciton annihilation (EEA) in a similar SQAB dimer, which has a relaxation time of about \qty{30}{fs} \cite{maly_signatures_2020,maly_coherently_2020-2}. 

Similar observations can be made in the experimental data at the 2Q spectral position (Figure~\ref{fig3_Exp_vs_Sim}, third column). The $S^{(3)}$ signal is zero within the experimental error, at the 2Q position, which is expected since $S^{(5)}$ is the lowest order that can contribute at the 2Q position. The 1Q and 2Q $S^{(5)}$ signals have peaks at the same $\omega_t$. The 2Q signal, which in principle could display emission and absorption at new spectral locations, displays only a single emission band and a single absorption band along $\hbar\omega_t$, just as the 1Q $S^{(5)}$ signal, again consistent with the system being mostly relaxed to the lowest-energy single exciton, even after double excitation. Similarly to the signal at the 1Q position, the sign of the signal flips on moving to $S^{(7)}$ and again to $S^{(9)}$. Unlike in the 1Q spectra, for the higher orders ($S^{(7)}$ and $S^{(9)}$), the signal near $\hbar\omega_{t}=\qty{1.72}{eV}$ is the most prominent, as the lower-energy feature is noisier.

We perform a more detailed analysis of the spectra by considering the long-$T$ limit in which the system has fully relaxed to $\ket{i}$, regardless of the initial excitation number. We simulated spectra using the Ultrafast Spectroscopy Suite (UFSS) \cite{rose_efficient_2021, rose_automatic_2021}. By learning the set of parameters in the model that are consistent with the linear and 2D spectra, we reveal the underlying physics of the dSQBC dimer. We demonstrate below how we can recover some of the information about the higher-lying excited states in the long-population-time limit by performing a detailed comparison of $S^{(3)}$ and $S^{(5)}$, rather than analyzing $S^{(3)}$ at smaller population times. That comparison works even when the system has reached quasi-equilibrium in the lowest-energy singly excited state, removing the need for detailed modeling of system dynamics.

Using UFSS, we simulated all the nonlinear orders observed experimentally at the 1Q and 2Q positions. We used a modified version of the model originally developed for the SQAB system \cite{maly_wavelike_2020}. The model consists of two coupled monomers where each monomer consists of a three-level electronic system as described in more detail in Supporting Information Section~S1. Since the linear spectrum shows clear vibronic states, we modified the model to explicitly include two vibrational modes, one per site, in the Hamiltonian, whereas Ref.~\cite{maly_wavelike_2020} treated all vibrations as a Markovian bath. We include the lowest six vibrational levels from the two vibrational modes, but we continue to treat the other modes as a Markovian bath. The pump spectrum strongly affects the peak heights in the 2D spectra, so obtaining quantitative agreement between simulations and experiment requires that these effects be included, which is a standard feature of UFSS \cite{rose_efficient_2021}. Therefore, all nonlinear signals were simulated using chirped pump pulses chosen to match the experimental intensity spectrum. We fit the experimental pulse intensity spectrum $I(\omega)$, found through a linear spectrally resolved intensity measurement, to the sum of five Gaussians and set $\tilde{\epsilon}(\omega)$ to the square root of the result. To quantify the chirp, we extract the group-delay dispersion (GDD) by fitting the spectral phase data retrieved through frequency-resolved optical gating (FROG) with a second-order polynomial. From the fit, we obtain a value of $\pm$\qty{36}{fs^2} for the GDD. The sign is not known directly in second-harmonic-generation FROG. We chose the positive value for our simulations as described in more detail in Supporting Information Section~S10. Because the experimental results show that the system was mostly relaxed to the lowest-energy exciton before the probe arrived, we cannot determine the relaxation rates. Instead, simulations were performed at a fictitious population time large enough that all relaxation processes other than fluorescence (lifetime approximately \qty{1.6}{ns} \cite{maly_wavelike_2020}) are complete. To determine the set of parameters in the model best describing our investigated sample dSQBC, we match simulated spectra to experimental spectra in three steps. First, we fix most of the static Hamiltonian and bath coupling parameters by reproducing the linear absorption spectrum. In the second step, we establish the transition dipole moment between exciton $\ket{i}$ and the biexciton $\ket{ij}$, as well as the biexciton binding energy, by matching the intensity ratio of the GSB and ESA regions in the third-order 1Q spectrum. In the last step the remaining unknown transition dipole moment is fixed by comparison of relative peak intensities between $S^{(3)}$ and $S^{(3)}$ signals. Steps one and two are described in more detail in Supporting Information Section~S1 and step three in Supporting Information Section~S6. No further fitting was performed. Consequently, the simulated seventh- and ninth-order spectra, as well as other features in the third- and fifth-order spectra constitute predictions of the model and are compared qualitatively with experiment without additional parameter adjustment. In Figure~\ref{fig3_Exp_vs_Sim}, columns two and four show the simulated nonlinear orders. The computational orders are scaled so that the maximum of $S^{(2n+1)}$ is consistent between computation and experiment.

\begin{figure}[tp]
\centering
\includegraphics[width=0.9\textwidth]{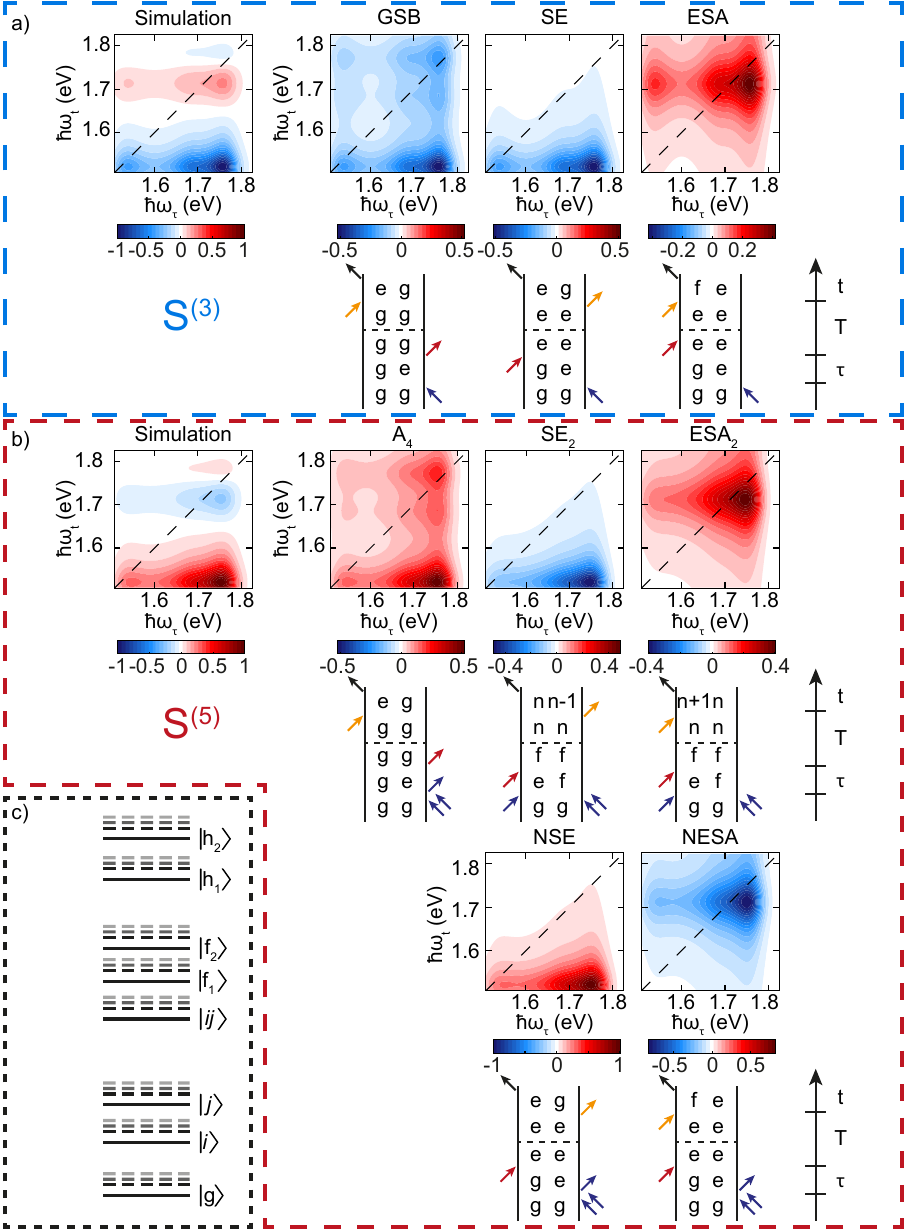}
\caption{Separated simulated signal contributions and examples of corresponding double-sided Feynman diagrams. a)~$S^{(3)}$ signal and GSB, SE, and ESA signal contributions with exemplary corresponding double-sided Feynman diagrams. b)~$S^{(5)}$ signal and A$_{(4)}$, SE$_{(2)}$, ESA$_{(2)}$, NSE and NESA signal contributions with exemplary corresponding double-sided Feynman diagrams. c) Energy scheme of the two coupled three-level systems representing the two monomer units.
}\label{fig4_SepSignalContrib}
\end{figure}

The simulations and experiments are in excellent agreement. Along the detection-frequency axis, we see the same two bands, one indicating GSB and SE from $\ket{i}$, the other indicating ESA from $\ket{i}$, as in the experiment. Further GSB peaks are overlapped by the ESA signal and are not distinguishable. For the higher orders, the locations of the peaks remain the same as in the lowest order, with the sign being inverted every second order, as in the experiment. This alternation is guaranteed in our simulations, since we have deliberately chosen a long population time in which the population is only in $\ket{i}$ when the probe arrives. The excellent agreement between the simulated and experimental peak heights could only be achieved by including the effects caused by the spectral shape of the electric field of the pump pulses. The excellent agreement in Figure~\ref{fig3_Exp_vs_Sim} between experiment and simulation demonstrates, on the one hand, that our model captures the underlying excitonic structure and processes contributing to the various nonlinear orders and, on the other hand, that the experimental separation into nonlinear orders now opens the door for quantitative evaluations because artifacts from higher orders are removed.

%
The UFSS algorithm allows us to separate our simulated spectra into the underlying Liouville pathways, revealing which specific signal contribution is present for each peak at the various nonlinear orders. Figures~\ref{fig4_SepSignalContrib}a and b show the simulated $S^{(3)}$ and $S^{(5)}$ 2D maps, respectively. The $S^{(3)}$ signal is divided into the usual contributions of GSB, SE, and ESA. The GSB signal is the negation of the absorption A, and the SE and ESA originate from singly excited states produced by pump excitation \cite{rose_interpretations_2023}. The $S^{(5)}$ signal can similarly be broken into five contributions: there are SE and ESA processes originating from doubly excited states produced by the pump, which we call SE$_2$ and ESA$_2$ \cite{rose_interpretations_2023}, and three ``negations'' of the processes contributing to $S^{(3)}$. These negations are processes in which the system is in the same state after the pump pulses as in the analogous lower-order contributions, and the probe then interacts with the same state after the last pump interaction, but the sign is inverted with respect to $S^{(3)}$, analogous to the case of normal GSB that negates absorption. Here, for $S^{(5)}$, we call the negated contributions A$_{(4)}$ -- since the negation of the GSB is a positive contribution to absorption --, negated SE (NSE), and negated ESA (NESA). The total $S^{(2n+1)}$ signal is normalized, so its maximum value is 1, and the contributions of the shown pathways sum up to the total. 

Below each respective 2D map, we include an exemplary double-sided Feynman diagram. In these Feynman diagrams, $\ket{g}$, $\ket{e}$, and $\ket{f}$ correspond to the ground state, a singly excited state, and a doubly excited state, respectively. The state $\ket{n}$ can represent either a singly or a doubly excited state, depending on the completion of the exciton--exciton annihilation during the population time. All symbols on the left of each diagram represent ket vectors, while all symbols on the right represent bra vectors. In the diagrams, blue arrows indicate an interaction with the first pump pulse, red ones with the second pump pulse and orange ones with the probe pulse. Furthermore, $\ket{n+1}$ corresponds to $\ket{n}$ being excited, while $\ket{n-1}$ represents a state originating from a de-excitation of $\ket{n}$. The scheme in Figure~\ref{fig4_SepSignalContrib}c shows the excitonic states that result from the coupling of two monomers, each of which is set up as an electronic three-level system. In addition, the electronic states are coupled to two vibrational modes with six vibrational levels each, which are depicted as dashed lines. For better readability, not all vibronic states are included in the scheme.

We start by discussing what we can learn about the system when considering only the lowest-order spectrum $S^{(3)}$. The clear presence of the positive ESA band allows us to extract the biexciton binding energy and the dipole strength that connect the lowest-energy exciton to the biexciton. Extracting the biexciton binding energy is somewhat complicated because the $\ket{j}$ exciton is a vibronic doublet, so we cannot determine the bare biexciton energy simply by adding the excitation energies to $\ket{i}$ and $\ket{j}$. In addition, the GSB band for the $\ket{j}$ doublet spectrally overlaps with the ESA band. Our model includes a coupling $K$ between the three doubly excited electronic states; $K$ shifts the biexciton energy compared to independent excitations, which is the origin of biexciton binding in the model. We determine $K$ from the position $\omega_t$ of the ESA band in $S^{(3)}$ and define the biexciton binding energy as the energy difference between the biexciton energy and the ``bare'' value obtained with $K=0$. The ESA band at $\hbar\omega_t=\qty{1.71}{eV}$ shows the energy separation between $\ket{i}$ and $\ket{ij}$. Using our model, we find that the binding energy of the $\ket{ij}$ biexciton is about \qty{0.05}{eV}, and that the transition dipole is $\mu_{ij,i}\approx\mu_{i,g}$, where $g$ indicates the electronic ground state.

Since the excited populations in the simulations have relaxed to the exciton state $\ket{i}$ by the time the probe arrives, there are no clear spectral features related to the exciton $\ket{j}$ found in the simulated spectra; similarly, there are no signatures of the doubly excited states of higher energy because the corresponding ESA transitions from $\ket{i}$ are outside the spectral window of the probe. However, even though the system has fully relaxed to $\ket{i}$ when the probe arrives, there are still spectral signatures of the dipole strength connecting the exciton $\ket{j}$ to the doubly excited electronic states, which we can use to extract that ratio in comparison with experiment. The key feature of the 1Q $S^{(5)}$ signal that we exploit is that the peaks along the $\omega_\tau$ axis occur in the same frequency regions as the $S^{(3)}$ signal, but have different relative weights. The weight change occurs because the $S^{(5)}$ signal has contributions from the Liouville pathways that include two additional pump-induced transitions compared to the $S^{(3)}$ signal. For a subset of diagrams, these transitions occur between the singly and doubly excited states, and later relaxation of the population does not remove these effects. These diagrams are therefore imprinted with information about the dipole strength of various singly to doubly excited states. From the change in peak weights with $\omega_\tau$ in $S^{(3)}$ and $S^{(5)}$ we can extract information on the pump-weighted average dipole strength,
\begin{equation}
d_{e} = \sqrt{\sum_f\mu_{fe}^2 I_\text{pump}(\omega_{fe})/\sum_f I_\text{pump}(\omega_{fe})},
\label{eq_de_weighted_dipole_Strength}\end{equation}
where $e$ indicates a singly excited state (either $\ket{i}$  or $\ket{j}$), $f$ indicates a doubly excited state, $\mu_{fe}$ is the dipole strength between them and $I_\text{pump}(\omega_{fe})$ is the pump intensity at the frequency difference between the two states.

To validate the agreement between simulation and experiment, we compared the ratio of peak intensities in two regions of $\omega_t$: the GSB- and SE-dominated signal for $\hbar\omega_t$ between \qty{1.51}{eV} and \qty{1.60}{eV} and the ESA signal for $\hbar\omega_t$  between \qty{1.67}{eV} and \qty{1.75}{eV}. We integrated the signals between $\hbar\omega_\tau = \qty{1.50}{eV}$ and \qty{1.62}{eV} for the peak corresponding to exciton $\ket{i}$ and between $\hbar\omega_\tau = \qty{1.62}{eV}$ and \qty{1.82}{eV} for the peak corresponding to exciton $\ket{j}$. The regions of interest are shown in Supporting Information Figure~S10. The ratios were then calculated as 
\begin{equation}
 R_{m} =\frac{\text{Peak}_{i,S^{(m)}}}{\text{Peak}_{j,S^{(m)}}},\label{eqPeakRatiosS2}
\end{equation} 
for $m=3$ and $m=5$. Our observable is the ratio of these ratios,
\begin{equation}
 X =\frac{R_{(3)}}{R_{(5)}},\label{eqPeakRatiosS2S4}
\end{equation}
which quantifies the difference in these normalized peak amplitudes between the $S^{(3)}$ and $S^{(5)}$ signals.
We performed this analysis for both the ESA signal and the overlapping GSB and SE signals, giving us the two values $X_\text{ESA}$ and $X_\text{GSB,SE}$, respectively. 

We chose this procedure since the intramanifold relaxation of excitons is nearly complete at our measured population time, so cuts along $\omega_t$ at fixed $\omega_\tau$ have approximately the same shape, with only the overall amplitude varying, as shown in Supporting Information Figure~S9. Simulations at shorter $T$ show that such cuts vary along $\omega_t$, but those variations disappear when the system has relaxed to the $\ket{i}$ state before the probe arrives, which corresponds to our experimental condition. The amplitudes of these peaks at different $\omega_\tau$ still contain information from excitation, which in the $S^{(5)}$ signals includes pathways that passed through doubly excited states, allowing us to determine $d_j/d_i$.

The ratio $X$ is only an indirect measure of the ratio $d_j/d_i$, because there are many other pathways that contribute to the signal but do not depend on $d_i$ or $d_j$. Experimentally, we find $X_{\mathrm{ESA}} = 1.2$ and $X_{\mathrm{GSB,SE}} = 1.3$. If the system were fully relaxed, these ratios would be identical, because both signals arise from a single excited state; these values being nearly equal is consistent with relaxation to the $\ket{i}$ exciton being nearly complete at $T=\qty{100}{fs}$. In Supporting Information Section~S11, we explore how residual $\ket{j}$ exciton population affects the $X$ ratios. To compare $X$ with simulation, we must find a set of parameters that is consistent with the linear absorption and $S^{(3)}$ spectra and then vary $d_i$, $d_j$ within these constraints. The transition dipoles from the ground to the singly excited states are fixed by the linear absorption spectrum, and the $S^{(3)}$ spectrum fixes $\mu_{ij,i}$, leaving one free transition dipole parameter in our model. We vary this free parameter and simulate both $S^{(3)}$ and $S^{(5)}$ to determine the ratio $X_{\mathrm{GSB,SE}}$, which is identical to $X_{\mathrm{ESA}}$ in the simulation due to the long population time limit, as shown in Supporting Information Section~S6 in more detail. By matching the average experimental value of $X=1.25$, we find that $d_j\approx 2d_i$. Since the experimental $X$ ranges from $X_\text{ESA}=1.2$ to $X_\text{GSB,SE}=1.3$, we determine an error bound on $d_i/d_j$ of about 5\%. In addition, the ratio $d_j/d_i$ varies by about 10\% as we change the coupling $J$  between the two monomer units, which we cannot tightly constrain without dynamics. We discuss this uncertainty in Supporting Information Section~S1. With both sources of error, we can constrain $d_j/d_i$ to lie in the range $[1.6,2.0]$. Our simulations show that $X_{\mathrm{GSB,SE/ESA}}$ qualitatively follows $d_j/d_i$, indicating that we have identified the quantity on which $X$ reports. 

It is important to note that the ratio $X$ is affected by chirp, and therefore we include the experimentally measured group-delay dispersion in our simulations. Otherwise, there would be no way to disentangle the contributions of chirp from the contributions of the relative dipole moments. Chirp affects the relative weight of the peaks along the excitation axis for all orders above third. Chirp changes higher-order peak amplitudes because when chirp is present, some transitions are excited before others. The loss of ground-state population and the presence of selected single excitation populations, when the lagging frequencies arrive, change the strength of excitation at those lagging frequencies. The impact of chirp on 2DES has been explored previously \cite{binz_effects_2020,yamaya_effects_2025,tekavec_effects_2010}.

A similar analysis of $X$ could in principle be performed on the peak ratios of the seventh-order signal, as compared with the fifth order. Such an analysis would attempt to quantify the dipole strength connecting double-excited states to triple-excited states. By analyzing successively higher orders, one can in principle build up quantitative information about the dipole ladder of the system. However, the analysis of higher-order signals becomes increasingly challenging with increasing order because each signal contains not only new pathways including highly excited states, but also contributions from all pathways already present in the lower-order signals\cite{rose_interpretations_2023}. Therefore, as a consequence of the reduced signal-to-noise ratio of the extracted seventh- and ninth-order signals, we did not perform this extended analysis. For measurements with a better signal-to-noise ratio, this analysis might be possible.

The experimental $S^{(5)}$ 1Q line shapes are slightly different from the $S^{(3)}$ 1Q line shapes. In particular, the $S^{(3)}$ peaks at $\hbar\omega_\tau=\qty{1.76}{eV}$ shift to $\hbar\omega_\tau=\qty{1.74}{eV}$ in $S^{(5)}$. The simulations in Figure~\ref{fig3_Exp_vs_Sim} do not fully account for the peak shift and we are uncertain of its cause. The separated orders can be used to measure exciton–exciton annihilation dynamics as a signature of exciton transport. We have shown previously, using transient absorption, that the rise of the fifth-order signal with population time can be used to quantify exciton transport because for annihilation to occur, the excitons first must propagate to get into close spatial proximity\cite{maly_wavelike_2020}. In addition, the dynamics of the seventh and higher orders can reveal information on the annihilation probability. While in earlier work\cite{maly_separating_2023}, we used only the frequency-integrated time evolution, more details about the microscopic processes of exciton–exciton interaction are expected to be found in the higher orders of the 2D line shapes. In the present example of the dimer, the two excitations are essentially co-localized right after excitation. Thus, there is no additional time scale that would represent exciton diffusion, and thus we did not measure it here. However, in general, such observations are useful for more complex samples. Figure~\ref{fig4_SepSignalContrib} shows the contributions to the $S^{(3)}$ and $S^{(5)}$ signals from the various Liouville pathways. The NSE/NESA, SE$_2$, and ESA$_2$ pathways of $S^{(5)}$ show a new spectral peak at $\hbar\omega_\tau=\qty{1.72}{eV}$ corresponding to the energy difference between the $\ket{ij}$ biexciton and the $\ket{i}$ exciton energy. This additional peak causes a small but noticeable shift in the $\omega_\tau$ line shape of the $\ket{j}$ doublet, which is visible by comparing to the $S^{(3)}$ line shapes. However, in the full simulations of Figure~\ref{fig3_Exp_vs_Sim}, the shift in line shape is not as strong and originates from finite-pulse and re-weighting effects, because after EEA and relaxation to the $\ket{i}$ exciton, these new peaks exactly cancel between the NSE/NESA pathways and the SE$_2$/ESA$_2$ pathways. If there were still a biexciton population, the cancellation would not be complete, and a dynamic peak shift might be observed as a function of population time. Note that this new peak does not appear in the A$_{(4)}$ pathways, but A$_{(4)}$ shows a more subtle change in line shape, which is due to the re-weighting of pathways, which is central to our analysis of the ratio $X$ above.


In the present work, we have experimentally shown the extraction of nonlinear orders in two-dimensional electronic spectra up to the $S^{(9)}$ signal, corresponding to ninth order in perturbation theory in the limit of a single, weak probe-pulse interaction. The separation of orders was achieved for one-quantum and two-quantum excitation positions through systematic intensity variation. Experiments were carried out on the squaraine dimer dSQBC. The extraction of the various orders revealed differences in the line shapes for the higher-order signals. Most significantly, we observed a shift of the 1Q peak position from the $S^{(3)}$ to the $S^{(5)}$ signals along the $\hbar\omega_\tau$ axis. Furthermore, a direct comparison of the $S^{(5)}$ signal at the 1Q and 2Q positions revealed significant differences in the line shape of the same-order signals: While the signal at the 1Q position consists of two peaks along $\hbar\omega_\tau$, the 2Q signal only shows one broad peak. 

All nonlinear order contributions were reproduced with simulations showing quantitative agreement. The separated $S^{(3)}$ and $S^{(5)}$ signals allowed us to quantify the strength of the dipole couplings to the doubly excited states. To gain information on the dipole transition strength, we compared the ratios of the integrated peak intensities of the $S^{(3)}$ and $S^{(5)}$ signals. Through the change in peak intensity we could match the simulation to the experiment and therefore describe the higher-excited states of our investigated sample. Our simulations also showed that the ratio of the pump-weighted average dipole strength of the two singly excited states qualitatively follows the ratio between SE and ESA signals ($X_\text{GSB,SE/ESA}$) proving that investigating $X$ presents one way to analyze the dipole strengths of the higher-excited state. Note that this  analysis of the dipole strength of higher-excited states was made possible by the order separation.

Furthermore, the simulations allowed us to split the signals into their underlying contributions which we described using double-sided Feynman diagrams. Through the analysis of the differences between the lower- and higher-order signals, we extracted information about line-shape changes caused by the interaction with the highly excited states. We inferred that the ESA$_{2}$ and SE$_{2}$ contributions only visible in the higher-order signal include interactions with the higher-excited states. In these signals, an energetic shift is visible, which provides insight into the energetic structure of the highly excited states.

In conclusion, the systematic extraction of several perturbative terms of artifact-free nonlinear-order signals in 2D electronic spectroscopy and the simulations using UFSS provide a unique spectroscopic framework to derive system properties, such as the energetic structure, coupling between states and transition dipole moments, also for higher-excited states. The quantitative agreement needed for this analysis is only achievable through the separation of orders and the removal of higher-order artifacts.

\section{Experimental}

The details of the experimental setup have been discussed elsewhere \cite{maly_separating_2023, krich_separating_2025, buttner_probing_2025}, and here we briefly review the main features. All measurements were performed using an Yb laser (Pharos, Light Conversion) as a light source. The pulses provided have a temporal length of around \qty{0.5}{ps} and a central wavelength of \qty{1030}{nm} at a repetition rate of \qty{50}{kHz}. The pulses were spectrally broadened using an optical parametric chirped pulse amplifier (Orpheus, Light Conversion) and then separated into pump and probe by a beam splitter (178892, Layertech GmbH), with 90\% of the intensity split into the pump beam and the remaining 10\% split into the probe beam. The probe beam was delayed by a motorized linear delay stage (M-IMS1000LM-S, Newport) to measure specific population times. Both pulses were compressed by individual prism compressors. The pump pulse was additionally compressed by an acousto-optic modulator (AOM) pulse shaper (Quickshape, Phasetech) at \qty{50}{kHz} repetition rate. The pulse shaper was also used to provide double pulses with variable delay and intensity. After compression, the probe pulse had a duration of \qty{25}{fs} (full width at half maximum of the intensity) and the pump pulse of \qty{14}{fs}. The pump and probe duration were measured using frequency-resolved optical gating \cite{trebino_frequency-resolved_2002}. Both beams were focused in the sample position, and the beam diameters were determined by fitting a beam-profile measurement with a CMOS camera (acA1280-60gm, Basler) with a Gaussian. The width parameters of this Gaussian fit, corresponding to the diameters at $e^{-1}$ of the maximum amplitude of the beam, show values of $d_x=\qty{86}{\micro m}$ and $d_y=\qty{76}{\micro m}$ for the probe and $d_x=\qty{134}{\micro m}$ and $d_y=\qty{344}{\micro m}$ for the pump were obtained. The relative polarization between the pump and probe pulses was set to the magic angle of \ang{54.7}. 

The dSQBC sample was dissolved in toluene and the concentration was chosen so that the absorption peak at \qty{1.77}{eV} showed an optical density of approximately \qty{0.3}{OD} for a sample thickness of \qty{1}{mm}.

As described above, we determined the optimal intensities for the separation of the nonlinear orders by first analyzing the population time dynamics in a TA measurement. Then, the saturation behavior of the TA signal was determined at a single population time. Afterwards, the noise level of the 2D measurement was identified. From the saturation behavior and noise level, the relative error was calculated, which was used to determine the optimal intensities for separating the nonlinear orders in the 2D measurement. Figure~\ref{fig2_MethodScheme} and the corresponding text describe the measurements in more detail.

To separate the four nonlinear orders, we measured four 2D maps at the intensities corresponding to pump pulse energies of \qty{1.74}{nJ}, \qty{7.35}{nJ}, \qty{12.9}{nJ}, and \qty{15}{nJ}. The pump pulse energies for the power-dependent TA experiment and the 2D experiments were measured with a PD300-BB-50mW (Ophir Spiricon) radiometric photodiode power sensor.  For the 2D experiments the pump pulse energies were measured for a single pump pulse.  For each 2D map, 251 steps were measured for coherence time $\tau$ with a step size of \qty{0.44}{fs}. The delay steps, the different pump intensities, and the chopping of every second laser shot were performed on a shot-to-shot basis by the AOM pulse shaper. To collect the data, the pump beam was blocked for every second shot and the probe spectrum was detected with a spectrometer (Spektrometer Acton 2156, Princeton Instruments) and a line camera (HS-Kamera Serie3030, Entwicklungsbüro Stresing). The probe spectrum with the pump present was then divided by the probe spectrum in the next shot with the pump blocked. After this division, the logarithm was taken and in the last step the $\Delta$OD values for the different steps of $\hbar\omega_t$, $\tau$, intensity were averaged over all measurements taken in the experiment. Afterwards, the data was mirrored along the $\tau$ axis around $\tau=\qty{0}{fs}$ with the exception of the $\tau=\qty{0}{fs}$ data point itself, so that the correct 0Q signal was displayed after Fourier transform. The data were then Fourier transformed and Eq.~\eqref{eqVdmMatrix} was used as described above to extract the nonlinear signal contributions.

\section*{Acknowledgements}

The authors thank Christoph Lambert, Michael Moos, and Maximilian Schreck for providing and synthesizing the sample. We acknowledge funding from the European Research Council (ERC) within the Advanced Grant IMPACTS [101141366] and from the Natural Sciences and Engineering Research Council of Canada (NSERC) [597081-24].





\printbibliography

\ifarXiv
            

    \pdfximage{\supplementfilename}
    \edef\mypages{1-\the\pdflastximagepages} 
    

\section*{Supplementary material}

# Supporting Information: Nonlinear order separation in two-dimensional electronic spectroscopy quantifies properties of higher-excited states

Katja Mayershofer[1], Peter A. Rose[2], Julian Lüttig[2], Luisa Brenneis[1], Simon Büttner[1], Jacob J. Krich*[2,3], and Tobias Brixner*[1,4]

[1]Institut für Physikalische und Theoretische Chemie, Universität Würzburg, Am Hubland 97074, Würzburg, Germany
[2]Department of Physics, University of Ottawa, 150 Louis-Pasteur Pvt, Ottawa ON K1N 6N5, Ontario, Canada
[3]Nexus for Quantum Technologies, University of Ottawa, 25 Templeton Street, Ottawa ON K1N 6N5, Ontario, Canada
[4]Center for Nanosystems Chemistry (CNC), Universität Würzburg, Am Hubland 97074, Würzburg, Germany

*Email: jkrich@uottawa.ca, tobias.brixner@uni-wuerzburg.de

## S1 Simulation parameters

We modeled the SQBC dimer as two coupled monomers. Each monomer is modeled as a three-level electronic system with a single vibrational degree of freedom coupled linearly to each electronic level. We write the Hamiltonian for one monomer as

$$H_\alpha = T_\alpha + V_g(q_\alpha)\left|g\right\rangle\left\langle g\right| + V_e(q_\alpha)\left|e\right\rangle\left\langle e\right| + V_f(q_\alpha)\left|f\right\rangle\left\langle f\right| \tag{S1}$$

where $\alpha = b, c$ indexes the monomer, $T_\alpha$ is the kinetic energy of the vibrational mode, and

$$V_g(q_\alpha) = \frac{1}{2}\omega_\alpha^2 q_\alpha^2, \tag{S2}$$

$$V_e(q_\alpha) = e_\alpha + \frac{1}{2}\omega_\alpha^2\left(q_\alpha - d_\alpha\right)^2, \tag{S3}$$

$$V_f(q_\alpha) = f_\alpha + \frac{1}{2}\omega_\alpha^2\left(q_\alpha - d_\alpha\right)^2, \tag{S4}$$

where $q_\alpha$ is the vibrational coordinate, $\omega_\alpha$ is the frequency of the vibrational mode, $d_\alpha$ is the displacement of the oscillator in the excited manifolds, and $e_\alpha, f_\alpha$ are the energy gaps between the ground state and the lowest-energy single-or double-excited states. Note that the Huang–Rhys factor $S_\alpha$ is found by dividing the reorganization energy by the energy of a phonon, $\hbar\omega_\alpha$, giving the formula $S_\alpha = \omega_\alpha d_\alpha^2/(2\hbar)$. To simulate the dimer, we include couplings $J, K, L$ that couple the single-, double-, and triple-excited states together. The total Hamiltonian can be written as

$$\begin{aligned} H = {} & T + V_{gg}\left|gg\right\rangle\left\langle gg\right| + V_{eg}\left|eg\right\rangle\left\langle eg\right| + V_{ge}\left|ge\right\rangle\left\langle ge\right| + J\left(\left|eg\right\rangle\left\langle ge\right| + \text{h.c.}\right) \\ & + V_{ee}\left|ee\right\rangle\left\langle ee\right| + V_{fg}\left|fg\right\rangle\left\langle fg\right| + V_{gf}\left|gf\right\rangle\left\langle gf\right| + K\left(\left|ee\right\rangle\left\langle fg\right| + \left|ee\right\rangle\left\langle gf\right| + \text{h.c.}\right) \\ & + V_{fe}\left|fe\right\rangle\left\langle fe\right| + V_{ef}\left|ef\right\rangle\left\langle ef\right| + L\left(\left|ef\right\rangle\left\langle fe\right| + \text{h.c.}\right) + V_{ff}\left|ff\right\rangle\left\langle ff\right|, \end{aligned} \tag{S5}$$

where $T$ is the total kinetic energy of the oscillators, $J, K, L$ are taken to be constants (not dependent on the vibrational degrees of freedom) and

$$V_{rs}(q_b, q_c) = V_r(q_b) + V_s(q_c). \tag{S6}$$

We model the light–matter interaction using the dipole and Condon approximations, where we have,

$$\mu_\alpha = \mu_{\alpha,e} |e\rangle \langle g| + \mu_{\alpha,f} |f\rangle \langle e| + \text{h.c.}, \tag{S7}$$

for a monomer and the total dipole operator is

$$\mu = \mu_b \otimes I_c + I_b \otimes \mu_c. \tag{S8}$$

where $I_\alpha$ are the identity operators for each monomer.

We simulate the effects of the bath by using Redfield theory [1]. We assume that the bath couples locally to each monomer unit, with the bath coupling being linear in the bath operator, so that the system–bath coupling is

$$H_{\text{SB},\alpha} = \sum_i c_{\alpha,e,i} |e\rangle\langle e| x_i + \sum_i c_{\alpha,f,i} |f\rangle\langle f| x_i \tag{S9}$$

for $\alpha = b, c$. The vibrations included in the system also couple to the bath via

$$H_{\text{SB,vib}} = \sum_i c_{\text{vib},i} q_\alpha x_i \tag{S10}$$

where we assume the vibration-bath coupling is the same for both vibrational modes and is also the same regardless of the electronic excitation. We take the continuum limit, where the system–bath coupling is treated in the Markovian limit as an ohmic spectral density with a Lorentz-Drude cut-off

$$J_s(\omega) = 2\omega\nu \frac{\gamma}{\omega^2 + \gamma^2}, \tag{S11}$$

where $\nu$ is the system–bath coupling strength and $\gamma$ is the cut-off frequency. Temperature effects are included by defining the real part of the spectral correlation function as

$$\Re\left[C(\omega)\right] = \frac{1}{2}\hbar J_s(\omega) \coth\left(\frac{\hbar\omega}{2k_B T}\right), \tag{S12}$$

and we do not include the imaginary part of $C(\omega)$, which provides the correct energy shifts for long times but gives the wrong energies for short times, which are relevant for linear absorption and for the excitation axis of ultrafast spectroscopies. This choice means that we can properly match most of the spectral features of interest, with the exception of the stimulated emission peak location along the detection frequency, which is red-shifted due to the reorganization of the bath. The mismatch can be seen in Figure S6a. We performed the simulations at room temperature, taking $k_B T = 0.025$ eV.

Finally, we model the exciton–exciton annihilation process, as well as the much slower fluorescence process, via a phenomenological rate $\gamma_r$, where the white-noise spectral density is represented as

$$J_s(\omega) = \begin{cases} \gamma_r & \omega > 0 \\ 0 & \omega < 0, \end{cases} \tag{S13}$$

where $\gamma_r = 1/\tau$, and $\tau$ is the lifetime of the associated process. The system–bath coupling is, in this case,

$$H_{\text{SB},\alpha} = \sum_i c_{\alpha,e,i} |e\rangle\langle f| x_i + \sum_i c_{\alpha,f,i} |g\rangle\langle e| x_i + \text{h.c.}, \tag{S14}$$

where h.c. describes the Hermitiona conjugate.

The parameters used in our dimer model system that represent the dSQBC molecule are shown in Tab. S1. We took the parameters given in the supplementary information of [2] for the SQAB system as a starting

Table S1: Parameters for the theoretical model of the dimer system

| **Quantity** | **Symbol** | **Value** |
|---|---|---|
| Singly excited energies | $e_c$, $e_b$ | $12\,310\,\mathrm{cm}^{-1}$, $14\,250\,\mathrm{cm}^{-1}$ |
| Doubly excited energies | $f_c$, $f_b$ | $26\,535.6\,\mathrm{cm}^{-1}$, $26\,535.6\,\mathrm{cm}^{-1}$ |
| Ground to singly excited transition dipole | $\mu_c$, $\mu_b$ | 1, 0.915 |
| Singly to doubly excited transition dipole | $\mu_{c,f}$, $\mu_{b,f}$ | 0.19, 4.41 |
| One-exciton couplings | $J$ | $-300\,\mathrm{cm}^{-1}$ |
| Two-exciton couplings | $K$ | $-354.9\,\mathrm{cm}^{-1}$ |
| Three-exciton couplings | $L$ | $-300\,\mathrm{cm}^{-1}$ |
| Huang–Rhys factors of singly excited states | $S_{c,e}$, $S_{b,e}$ | 0.23, 0.13 |
| Huang–Rhys factors of doubly excited states | $S_{c,f}$, $S_{b,f}$ | 0.23, 0.13 |
| Vibrational energies | $\hbar\omega_{c,g}$, $\hbar\omega_{b,g}$ | $1300\,\mathrm{cm}^{-1}$, $1280\,\mathrm{cm}^{-1}$ |
| Site-bath coupling of singly excited states | $\nu_{c,e}$, $\nu_{b,e}$ | $385\,\mathrm{cm}^{-1}$, $430.5\,\mathrm{cm}^{-1}$ |
| Site-bath coupling of doubly excited states | $\nu_{c,f}$, $\nu_{b,f}$ | $1400\,\mathrm{cm}^{-1}$, $1400\,\mathrm{cm}^{-1}$ |
| Vibration–bath coupling | $\nu_{\mathrm{vib}}$ | $350\,\mathrm{cm}^{-1}$ |
| Vibration- and site-cut-off frequency | $\gamma$ | $500\,\mathrm{cm}^{-1}$ |
| Internal conversion time | $\tau_{\mathrm{IC}}$ | 30 fs |
| Exciton relaxation time | $\tau_{\mathrm{FL}}$ | 1641.1 ps |

point, although we have used significantly larger values of the bath couplings because we do not include inhomogeneous broadening in our calculations and therefore need larger homogeneous line widths.

For the parameters determining the linear absorption (LA) spectra, we modified the parameters from [2] as necessary to match the LA spectra of each individual monomer unit. We then matched the LA spectrum of the dimer, which required further modifications of the parameters. For example, the Huang–Rhys factors of the vibrational modes were each reduced by about 20% to match the dimer.

For a chosen value of $J$, the parameters $e_c, e_b, \mu_c, \mu_b, S_{c,e}, S_{b,e}$ are well-determined by the linear absorption spectrum. We find that we can fit the dimer's linear absorption spectrum with values of $J$ ranging from $300\,\mathrm{cm}^{-1}$ to $-650\,\mathrm{cm}^{-1}$. We have chosen $J = -300\,\mathrm{cm}^{-1}$ as our primary choice because we find that it allows us to keep the parameters $e_c, e_b, \mu_c, \mu_b, S_{c,e}, S_{b,e}$ more similar to the values we find when matching the monomer spectra. However, we do not take these parameters to be definitive.

Instead, we have studied how our reported values for the ratio $\mu_{ij,i}/\mu_{i,g}$, the biexciton binding energy, and the ratio $d_j/d_i$ depend on the choice of $J$. The variation of these parameters is shown in Tab. S2. The error on $d_j/d_i$ for each set of parameters is due to the uncertainty in the quantity $X$, which is different for the SE band and the ESA band (see Figure S8). Since the current data do not allow us to fully determine $J$, we report that $d_j/d_i$ is constrained to lie in the range $[1.6, 2.0]$. The value of $J$ has a large impact on the dynamics of the system, with small $|J|$ giving slower dynamics than large $J$. For example, near complete relaxation of the singly excited states occurs on the order of 100s of femtoseconds for $J = -300\,\mathrm{cm}^{-1}$. In contrast, near complete relaxation of the singly excited states occurs on the order of 100s of picoseconds for $J- = 10\,\mathrm{cm}^{-1}$. $J = -300\,\mathrm{cm}^{-1}$ is therefore a more likely candidate for this system. However, without dynamics we cannot fully constrain $J$, and leave the study of dynamics to future work.

Table S2: Variation of extracted physical quantities with changing $J$ for positive chirp ($\beta^{(2)} = +36$ fs$^2$/rad)

| Chirp | $J$ | $\mu_{ij,i}/\mu_{i,g}$ | Binding Energy | $d_j/d_i$ |
|---|---|---|---|---|
| 36 fs$^2$/rad | -10 | 1.01 | -0.056 | $1.7 \pm 0.1$ |
| | -100 | 1.07 | -0.056 | $1.75 \pm 0.1$ |
| | -200 | 1.10 | -0.056 | $1.85 \pm 0.1$ |
| | -300 | 1.07 | -0.057 | $1.95 \pm 0.1$ |
| -36 fs$^2$/rad | -10 | 1.01 | -0.056 | $1.7 \pm 0.1$ |
| | -100 | 1.07 | -0.056 | $1.75 \pm 0.1$ |
| | -200 | 1.10 | -0.056 | $1.95 \pm 0.1$ |
| | -300 | 1.07 | -0.057 | $2 \pm 0.1$ |

Somewhere between $J = -300\,\text{cm}^{-1}$ and $J = -400\,\text{cm}^{-1}$, and for all values of $J < -400\,\text{cm}^{-1}$, as well as for $J > 0$, the model can no longer simultaneously match $X$ and the $S^{(3)}$ spectrum. For these incompatible values of $J$, a value of $d_j/d_i \approx 2$ is still associated with $X \approx 1.25$. However, the parameters that produce this value of $d_j/d_i$ also create an ESA band in $S^{(3)}$ that is inconsistent with the experiment. This increased strength is due partly to a value of $\mu_{ij,i}$ that is too large and partly to transitions from $|i\rangle$ to higher-lying excited states that become more intense. For $0 < J| \lessapprox 300\,\text{cm}^{-1}$ these higher-energy transitions from $|i\rangle$ are too small to be seen. Our key finding is that the physically meaningful ratio $d_j/d_i \approx 2$, varying by a little more than 10% over the range of $J$ values that are consistent with linear and nonlinear spectra.

## S2 Early-time dynamics of dSQBC

As mentioned in the main text, we simulated the 2D spectra for a population time of $T = 1\,\text{ps}$. We chose such a late population time because the experimental results are consistent with fast relaxation processes that are largely being completed, as indicated by the early population times ($T$ up to 200 fs) of the experimental measurement of transient absorption (TA), seen in Fig S1a. The 2D measurements shown in the main text were taken for $T = 100\,\text{fs}$. To better visualize the dynamics, we show cuts at $\hbar\omega_t = 1.52\,\text{eV}$, $1.67\,\text{eV}$, $1.72\,\text{eV}$ and $1.77\,\text{eV}$ in Figure S1b. The cuts demonstrate that the most significant dynamics occur before $T = 100\,\text{fs}$. Looking closely at the cut of $\hbar\omega_t = 1.52\,\text{eV}$ (Figure S1b, blue curve), the signal shows a quick increase in amplitude, starting at 0 fs. The signal then stops increasing at a population time of 80 fs. This behavior indicates that the early-time dynamics of the dimer molecule are nearly complete, so we neglect them by simulating 2D spectra at $T = 1\,\text{ps}$.

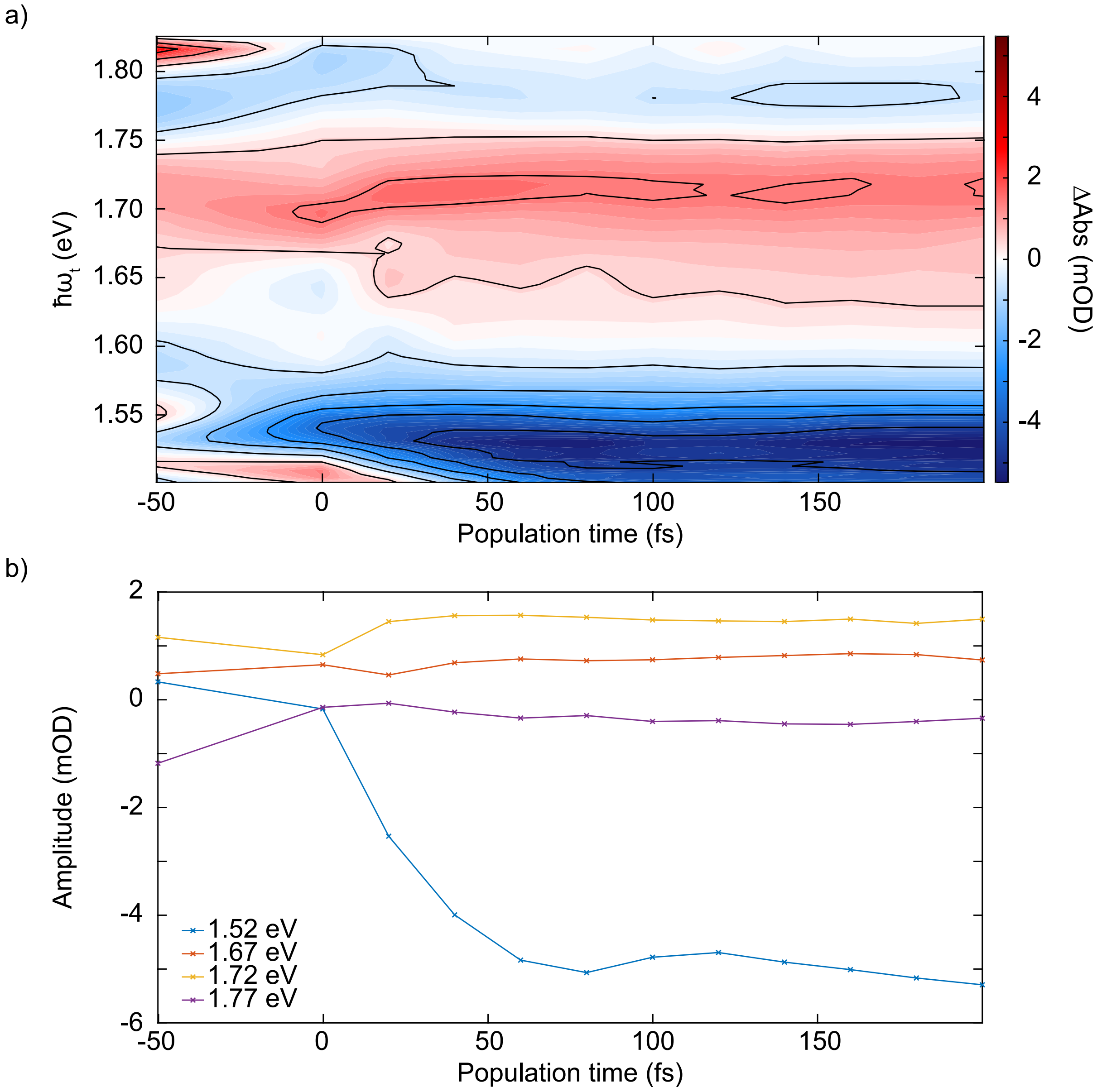


Figure S1: Early population time TA data for the dSQBC dimer. a) TA data for population times measured between $-50\,\text{fs}$ and 200 fs. b) Cuts taken from the TA data at $\hbar\omega_t = 1.52\,\text{eV}$, $1.67\,\text{eV}$, $1.72\,\text{eV}$ and $1.77\,\text{eV}$.

## S3 Saturation behavior of dSQBC

As described in the main text, an intensity-dependent TA measurement was performed to determine the saturation behavior of the investigated sample dSQBC. Figure S2a shows the intensity-dependent TA data at $\hbar\omega_t = 1.51\,\text{eV}$ and $T = 100\,\text{fs}$ and the exponential absorption fit described by Eq. (3) in the main text. Figure S2b shows the spectrally resolved $S_{\text{max}}$ (blue) with the corresponding standard errors (blue shaded area) with the linear absorption spectrum (orange) for reference. The $\hbar\omega_t$ value of 1.51 eV, at which the $S_{\text{max}}$ value was determined for the ideal powers, is indicated by the black dashed line. The value was chosen as it corresponds to the absolute maximal signal intensity along the $\hbar\omega_t$ axis in the 2D measurement of dSQBC. Between $\hbar\omega_t = 1.5\,\text{eV}$ and 1.65 eV, the $S_{\text{max}}(\omega_t)$ fit parameters follow the linear absorption spectrum, while the fit parameters deviate for larger $\hbar\omega_t$. This mismatch can be traced back to the overlap of the contributions of ESA, SE, and GSB between $\hbar\omega_t = 1.65\,\text{eV}$ and 1.72 eV. In the spectral region where these signals overlap, Eq. (3) does not provide a good fit, as seen from the large standard errors displayed in Figure S2. Figure S2c shows the spectrally resolved $I_{\text{sat}}(\omega_t)$ with the corresponding standard errors. Although $I_{\text{sat}}(\omega_t)$ varies mostly smoothly, there is a spike near $\hbar\omega_t = 1.62\,\text{eV}$, where the TA signal changes sign. There is a similar spike near $\hbar\omega_t = 1.75\,\text{eV}$, again where the TA signal changes sign. Above 1.75 eV, the fit quality is poor and large $I_{\text{sat}}$ are found.

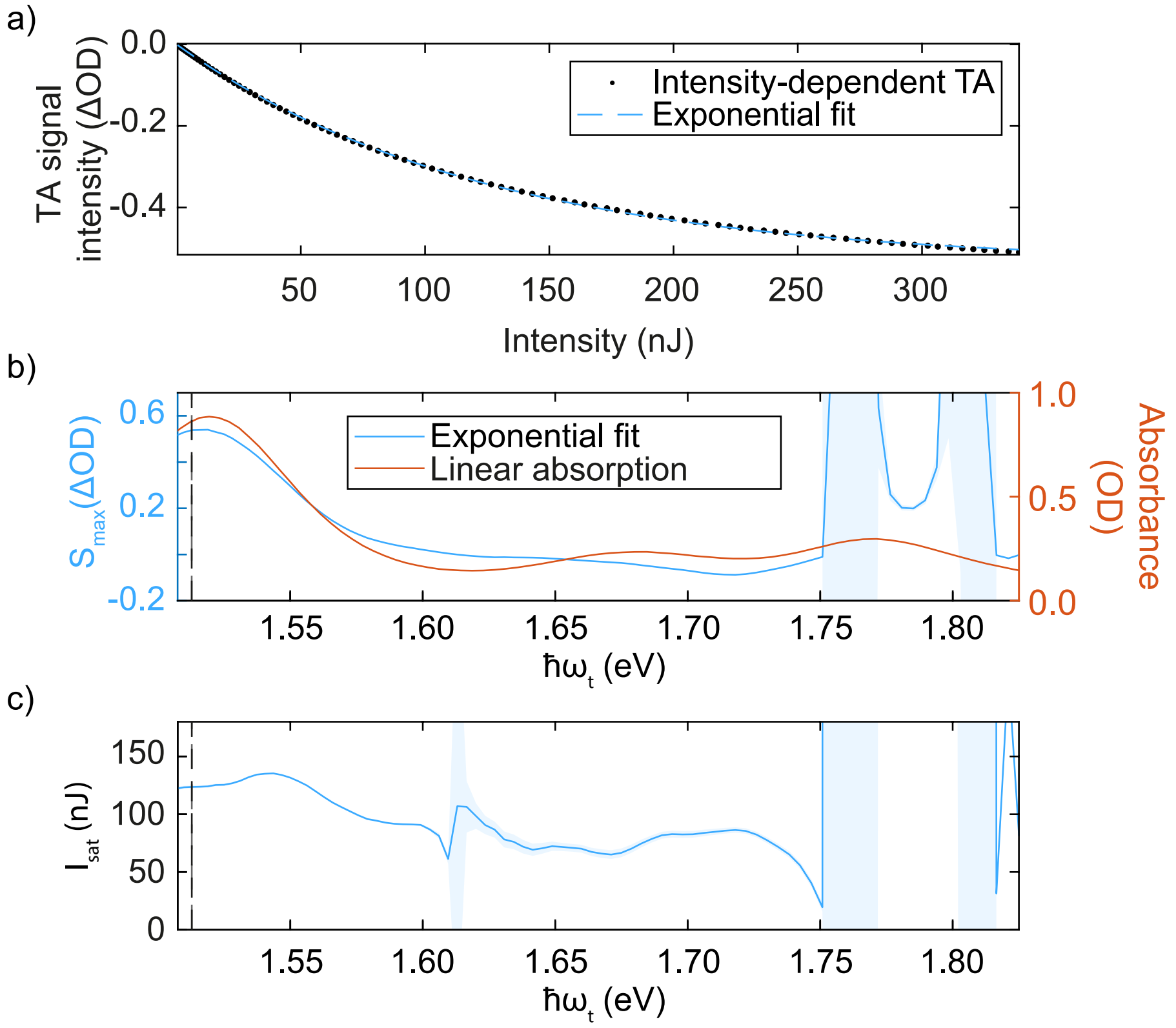


Figure S2: Intensity-dependent measured TA data and saturable absorption fits for the calculation of the ideal excitation intensities. a) Intensity-dependent TA data at a population time of $T = 100\,\text{fs}$. b) Spectrally resolved $S_{\text{max}}$ signal from the saturable absorption fit (blue line) with standard errors (blue shaded area) and linear absorption spectrum (orange). The value used for the calculation of the optimal excitation intensities is marked by the black dashed line. c) Spectrally resolved $I_{\text{sat}}$ values with standard errors (blue shaded area) from the saturable absorption fit of the TA data in a).

We show three cuts at frequencies where the fit does not work well in Figure S3. In two of the cuts, at $\hbar\omega_t = 1.62\,\text{eV}$ and $\hbar\omega_t = 1.75\,\text{eV}$ in Figure S3b and c, the data are nonmonotonic in intensity rather than showing clear saturation. Despite these non-saturating forms in some portions of the TA spectrum,

the excitation intensity values derived from Eq. (3) lead to an excellent separation of the nonlinear orders. Section S5 describes self-consistency checks, which show the quality of the separated orders.

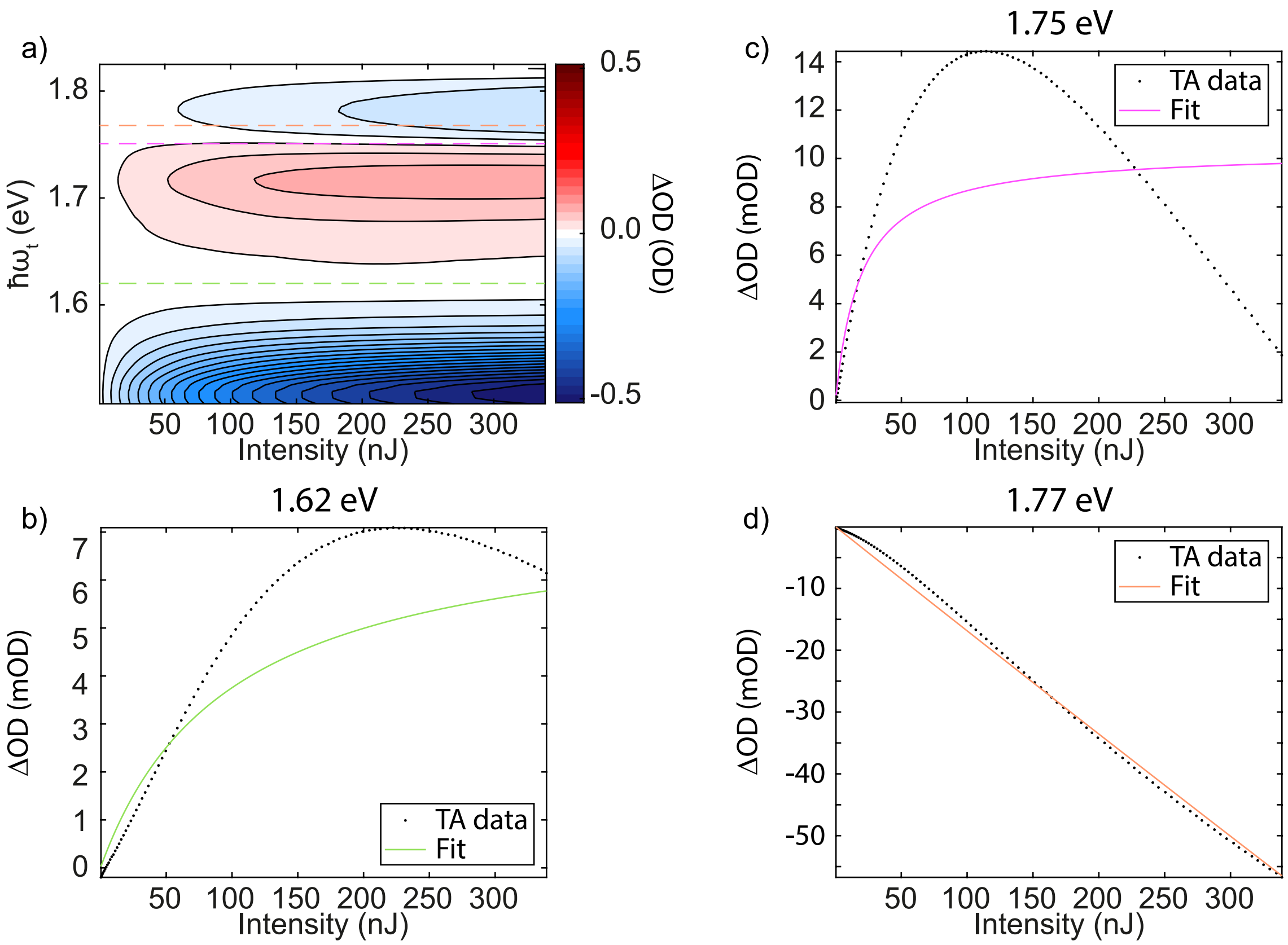


Figure S3: Spectrally resolved measured intensity-dependent TA data and cuts along with the corresponding fits. a) Spectrally resolved intensity-dependent TA data taken for a population time of $T = 100\,\mathrm{fs}$. b) Cut from the intensity-dependent TA data at 1.66 eV and corresponding fit. c) Cut from the intensity-dependent TA data at 1.75 eV and corresponding fit. d) Cut from the intensity-dependent TA data at 1.77 eV and corresponding fit.

# S4 Raw data

To extract the separated $S^{(2n+1)}$ signals shown in Figure 3 of the main manuscript, four measurements were taken at excitation intensities, i.e., pulse powers, of 1.74 nJ, 7.35 nJ, 12.9 nJ and 15 nJ. The 1Q and 2Q peak regions of these measurements are shown in Figure S4, showing similar peak structures for the 1Q and 2Q 2D spectra as in similar dimers [3] as well as theoretical simulations of dimers [4]. The 1Q region shows two peaks near the diagonal $\hbar\omega_t = \hbar\omega_\tau$: a negative peak at 1.52 eV and a positive peak at 1.77 eV. Both peaks are located slightly below the diagonal, i.e., to smaller values of $\hbar\omega_t$. Two major cross peaks between these diagonal peaks can be distinguished. The cross peak above the diagonal ($\hbar\omega_\tau = 1.52\,\mathrm{eV}$ and $\hbar\omega_t = 1.77\,\mathrm{eV}$) is positive while the cross peak below the diagonal ($\hbar\omega_\tau = 1.77\,\mathrm{eV}$ and $\hbar\omega_t = 1.52\,\mathrm{eV}$) is negative. Another smaller peak can also be seen at $\hbar\omega_t = 1.78\,\mathrm{eV}$ and $\hbar\omega_\tau = 1.77\,\mathrm{eV}$. In the 2Q region the signal also shows two main peak regions along $\hbar\omega_t$, analogous to the 1Q region. The peaks are also located at $\hbar\omega_t = 1.52\,\mathrm{eV}$ and $\hbar\omega_t = 1.72\,\mathrm{eV}$. However, only two broad main peaks can be distinguished. In contrast to the 1Q region the signals here show inverted signs, i.e., the main peak at $\hbar\omega_t = 1.52\,\mathrm{eV}$ is positive and the peak at $\hbar\omega_t = 1.72\,\mathrm{eV}$ is negative. This sign change reflects the different leading order of nonlinear response that is present at the 1Q and 2Q positions: corresponding to the change in the leading order signal from

the $S^{(3)}$ at the 1Q position to the $S^{(5)}$ at the 2Q position. Similar to the 1Q peaks, there is another smaller positive peak at $\hbar\omega_t = 1.78\,\mathrm{eV}$ and $\hbar\omega_\tau = 3.52\,\mathrm{eV}$.

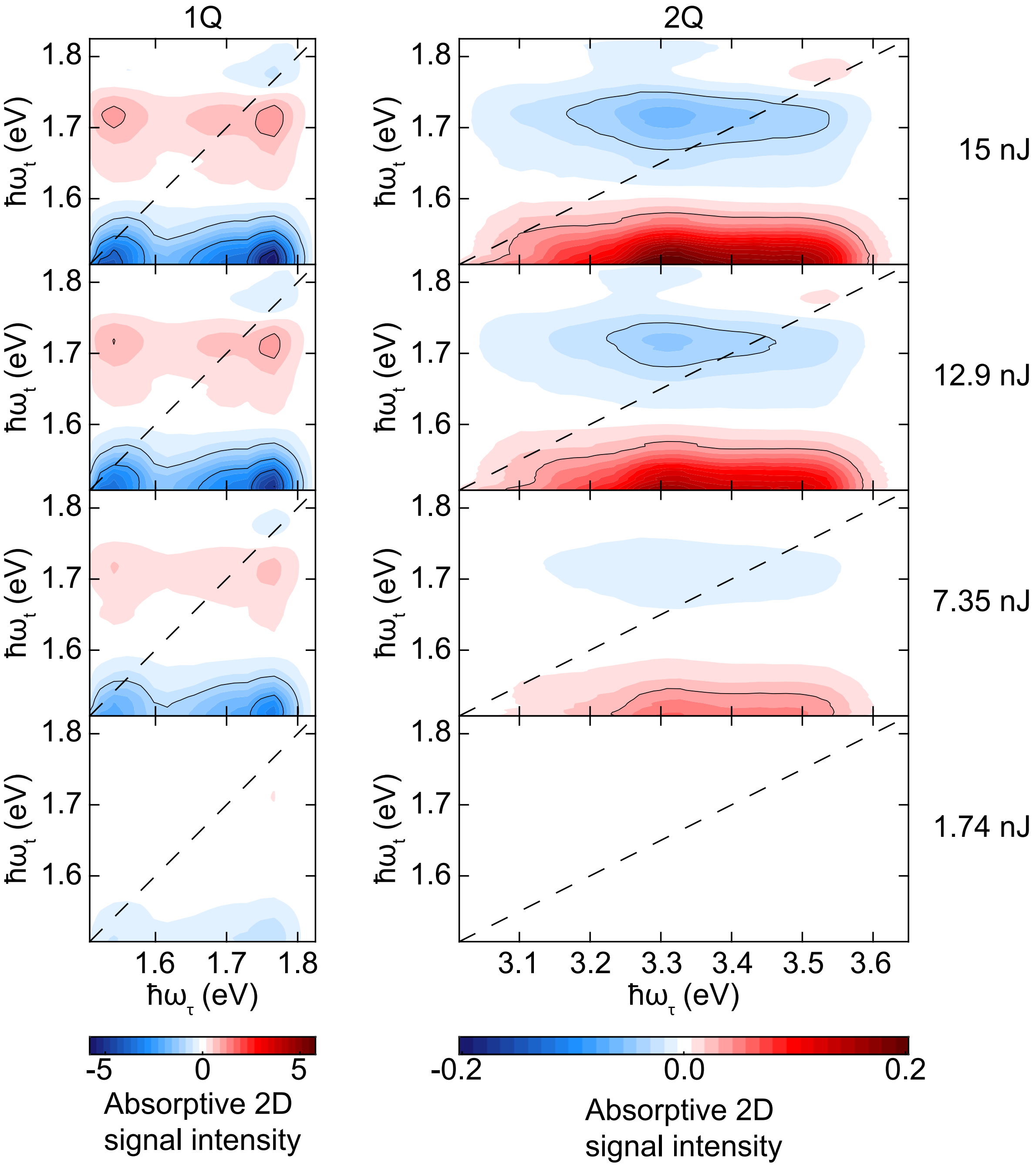


Figure S4: 2D measurements of dSQBC at excitation intensities 15 nJ, 12.9 nJ, 7.35 nJ, and 1.74 nJ.

## S5 Self-consistency checks of extracted orders

To verify that the extraction of the $S^{(2n+1)}$ signals worked correctly, we performed self-consistency checks by comparing the different order signals at the $r$Q spectral positions for $r = 0, 1, 2$, as described in [5]. The 0Q signal was integrated from $\hbar\omega_\tau = -0.83\,\text{eV}$ to $0.83\,\text{eV}$, the 1Q signal from $\hbar\omega_\tau = 0.85\,\text{eV}$ to $2.50\,\text{eV}$, and the 2Q signal from $\hbar\omega_\tau = 2.52\,\text{eV}$ to $4.17\,\text{eV}$. After integration, the three signals are scaled by a factor of $\frac{n}{\frac{1}{2}n-r}$, where $n$ corresponds to the order of the signal in pump pulse interactions and $r$ describes the spectral position. The integrated and scaled signals are shown in Figure S5. For the $S^{(2n+1)}$ signal, the scaled $r$Q signals should match for $0 \leq r \leq n$ and $r$Q should be zero for $r > n$ [5]. For the $S^{(3)}$ signal, the signals from the 0Q and 1Q position show perfect overlap, while the signal at the 2Q position is zero over the whole $\hbar\omega_t$ range, as expected. Comparing the $S^{(5)}$ signals at the three spectral positions also shows excellent agreement of the integrated signals. There are small differences in the scaled signals near $\hbar\omega_t = 1.52\,\text{eV}$, which we attribute to the increased noise level at the edges of the excitation and probe spectrum. These self-consistency checks show that we have obtained high-quality extracted orders even in the spectral region above $\hbar\omega_t = 1.75\,\text{eV}$ where the TA saturation fits were poor in Figure S2.

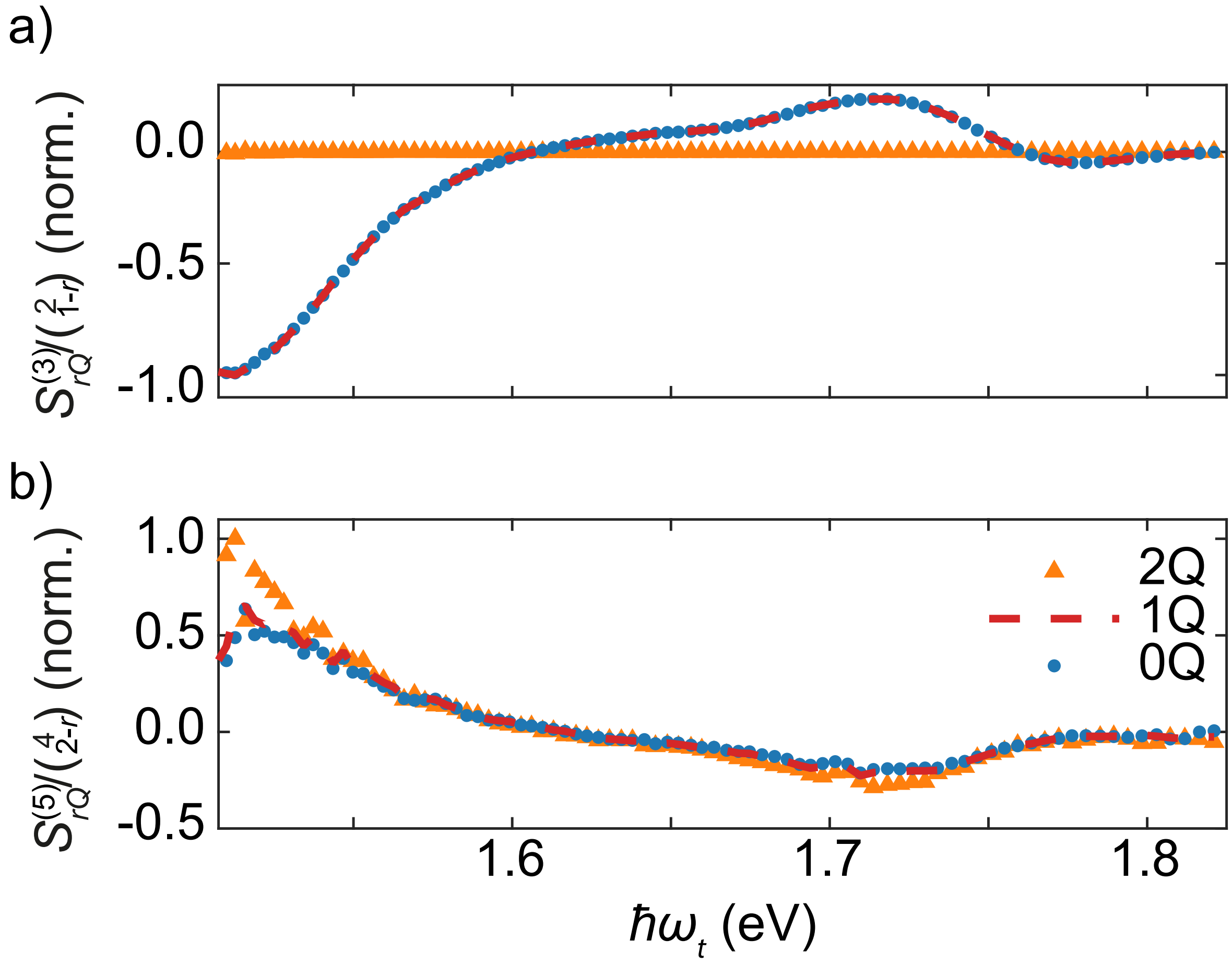


Figure S5: Self-consistency checks. a) Integrated 0Q, 1Q, and 2Q signals for $S^{(3)}$ signal. b) Integrated 0Q, 1Q, and 2Q signal for $S^{(5)}$ signal.

## S6 Constraining the simulation parameters

As discussed in the main text, we inferred the biexciton binding energy and the dipole transition strength in the exciton basis $\mu_{ij,i}$ by using our model to match the experimental $S^{(3)}$ 1Q signal. The transition dipoles between excitons vary both with state hybridization (controlled by $J$ and $K$) and with the site transition dipoles $\mu_{bf}$ and $\mu_{cf}$. We integrated the experimental and simulated results over $\omega_\tau$, normalized them separately, and plotted them together, as shown in Figure S6a and b. We then set $\mu_{bf} = 0$ and adjusted $K$ and $\mu_{cf}$ until we found a good match by eye. Adjusting $K$ moves the location of the ESA peak reporting on $|i\rangle \to |ij\rangle$, while adjusting $\mu_{cf}$ modifies $\mu_{ij,i}$, and thus the height of the ESA peak. We can vary the site couplings $\mu_{cf}$, $\mu_{bf}$ while keeping the excitonic coupling $\mu_{ij,i}$ unchanged, which defines a parametric curve in $(\mu_{cf}, \mu_{bf})$ space. We determine this curve by using a minimizer to find the values of $\mu_{bf}$ that produce the correct $\mu_{ij,i}$ for $\mu_{cf}$ in the interval $[0, 0.72]$, in steps of 0.01. Along this curve, the ratio $X$ described in the main text varies, and we find the point of best agreement. At the point of best agreement for $X$, the simulated $S^{(3)}$ ESA peak height no longer matches the experimental data, because higher-energy transitions from $|i\rangle$ are slightly enhanced for the best pair $(\mu_{cf}, \mu_{bf})$. In order to simultaneously match the linear, $S^{(3)}$, and $S^{(5)}$ spectra simultaneously, we must therefore iterate this procedure to self-consistency. We take $\mu_{bf}$ fixed, and then modify $\mu_{cf}$ to match $S^{(3)}$. This changes the value of $\mu_{ij,i}$ by about 1%, which re-defines the parametric curves. We then repeat the process of finding the best match to $X$. We find that a single additional iteration is sufficient. We now describe the determination of $K$, $\mu_{ij,i}$, and $X$ in more detail.

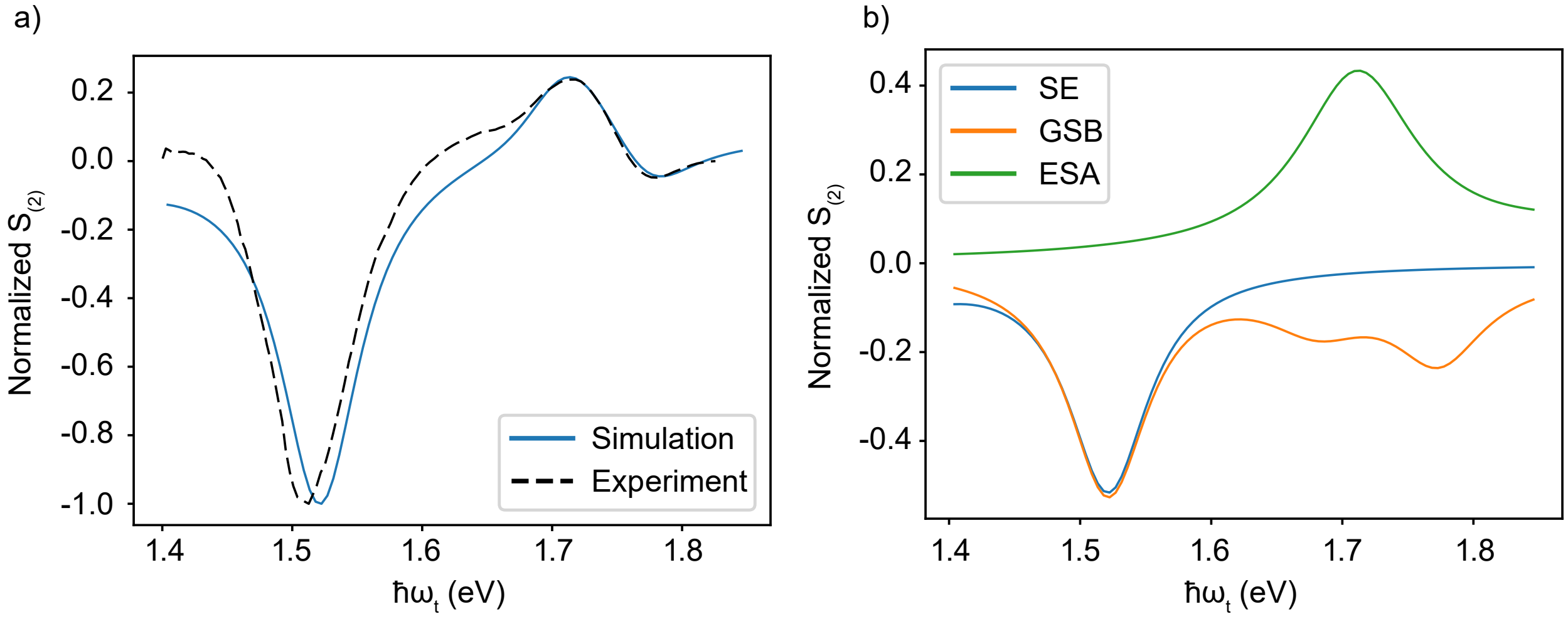


Figure S6: $S_{(2)}$ 1Q signal integrated along $\omega_\tau$. a) Comparison between the experimental and simulated signals. b) Shows the simulated GSB, SE and ESA signals separately, demonstrating the spectral overlap of the ESA signal corresponding to $i \to ij$ and the GSB signal for the $j$ doublet.

As mentioned in the main text, we cannot obtain either $K$ or $\mu_{ij,i}$ directly from the experimental data without using our model, since $|j\rangle$ is a vibrational doublet. If it were a single peak, we could define the bare biexciton energy to be the sum of the $|i\rangle$ and $|j\rangle$ peak locations and then find the binding energy by comparing the spectral position of the ESA peak to the expected value if the binding energy were 0. However, in our case it is unclear where exactly "$|j\rangle$" is, spectrally. Further complicating the analysis, the GSB signal has two peaks, due to the doublet, which appear with opposite sign and in the same spectral region as the ESA peak from $|i\rangle$ to $|ij\rangle$. This overlap can be seen in the simulated individual signal contributions in Figure S6b. The parameter $K$ couples the doubly excited states together and is responsible for the biexciton binding energy. We determine the biexciton binding energy by first finding the ESA peak location with $K = 0$. Then we find the peak location with the correct value of $K$ and subtract the bare value. We find a biexciton binding energy of about 50 meV.

We now consider the transition dipole moments. Just as with determining $K$, determining $\mu_{ij,i}$ requires a more complicated analysis than usual, since the ESA peak is located in the same spectral band as the GSB signal for the $j$ doublet. As discussed in the main text we can extract $\mu_{ij,i}$ and $\mu_{i,g}$ from the peak strength of the ESA signal in the $S^{(3)}$ spectrum and the linear absorption spectrum, respectively. We find that $\mu_{ij,i} \approx \mu_{i,g}$.

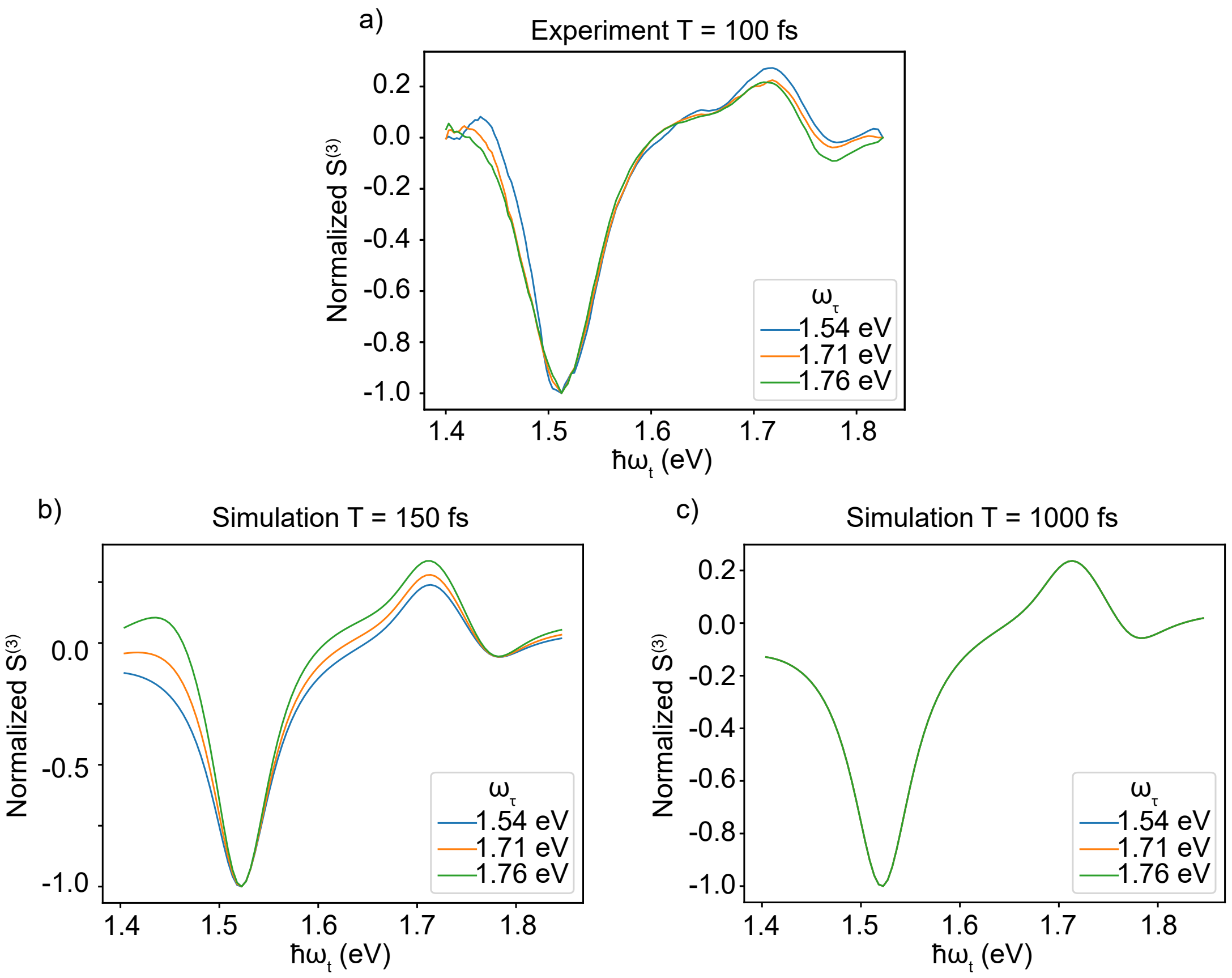


Figure S7: Cuts taken at different $\omega_\tau$ values for the experimental and simulated $S^{(3)}$ signal. a) Cuts along the $\hbar\omega_\tau$-axis for the experimental 1Q $S^{(3)}$. Cuts from the simulated $S^{(3)}$ signal with a coupling $J = 300\,\mathrm{cm}^{-1}$ at b) 150 fs and c) 1000 fs population time.

In the main text we discussed our system having mostly relaxed to the lowest singly excited state before the probe arrives. One indicator of this relaxation is that linecuts of $S^{(3)}$ have nearly the same shape with widely varying $\omega_\tau$, indicating that the interaction with the probe is the same independent of the initial excitation, as mentioned in the main text. Figure S7a shows three such cuts taken from the $S^{(3)}$ signal, showing good agreement in their line shapes after normalizing to have the same amplitude.

We compare these experimental line cuts to simulations at different population times in Figure S7b and c. Simulations at shorter population times, before relaxation in the singly excited manifold has significantly progressed, show clear differences in the line cuts. At later population times, including the $T = 1000\,\mathrm{fs}$ that we use in our other studies, the relaxation process is completed, and the cuts show matching shapes.

Since relaxation is mostly complete when the probe arrives at the experimental $T$, comparing the relative peak ratios between simulation and experiment is not influenced by additional system dynamics. We can now show the influence of the input parameters $\mu_{c,f}$ and $\mu_{b,f}$ on the ratios of the peaks. As described in the main text, we compared the ratios of the integrated peak intensities between the $S^{(3)}$ and the $S^{(5)}$ signal. The calculation steps are described by Eqs. (7) and (8). The integration ranges are shown in Figure S8a and b for the $S^{(3)}$ signal from the experiment and simulation, respectively. Figure S8c shows the results from Eq. (8) for different input parameters $\mu_{b,f}$ and $\mu_{c,f}$. The black line indicates the $(\mu_{b,f}, \mu_{c,f})$ contour

that matches the transition dipole strength $\mu_{ji,i}$, as described above . The star marks the best match to the mean of the experimental $X$ values for the ESA and SE peak regions. Figure S8d shows the $X$ values for the different ratios of the pump-averaged dipole strengths $d_j/d_i$ of exciton $|i\rangle$ and $|j\rangle$.

By matching the parameters extracted from the experiment, we can find the best parameters $\mu_{c,f} = 0.19$ and $\mu_{b,f} = 4.41$, which are tightly constrained by the $S^{(3)}$ and $S^{(5)}$ spectra.

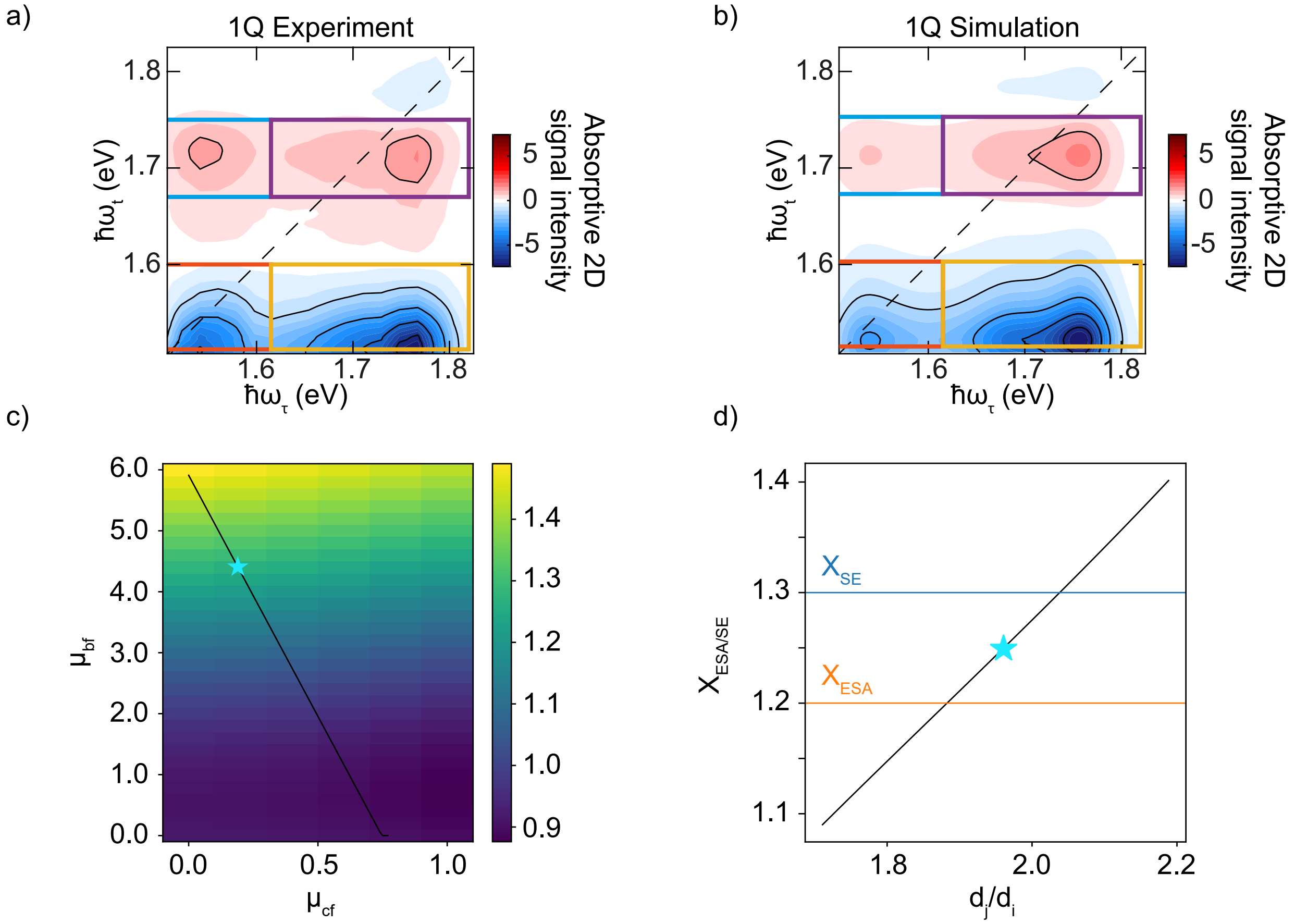


Figure S8: Regions of interest marked in the a) experimental and b) simulated 2D maps. c) Calculated ratio $X$ for different simulation parameters $\mu_{bf}$ and $\mu_{cf}$. d) Calculated ratio $X$ for ratios of the pump-averaged dipole strengths for the exciton $i$ and $j$. The black line in c) and d) represents the transition dipole strength $\mu_{ij,i} = 1.07$. The star marks the best match between simulation and experiment.

# S7 Cancellation of $ESA_2$ and $SE_2$ pathways by NESA and NSE pathways

As described in the main text, there are NESA and NSE pathways, which cancel with the $ESA_2$ and $SE_2$ pathways after exciton–exciton annihilation (EEA) is complete. Figures S9a–d show a subset of the Feynman diagrams describing the $ESA_2$ (a), NESA (b), $SE_2$ (c) and NSE (d) pathways. The $ESA_2$ and $SE_2$ (Figure S9a and c, respectively) diagrams show an EEA process during the population time. Comparison of these diagrams shows that the dipole couplings for each of the NESA diagrams are the same as for the $ESA_2$ diagrams, after EEA has completed, and similarly for the $SE_2$ and NSE. When summing over all diagrams, Figure S9e and f shows the cancellation of 99.99% of the peak amplitude, with the remainder likely due to residual numerical error.

The diagrams displayed in Figure S9 all involve three interactions with the first pulse and one interaction with the second pulse. The diagrams that involve one interaction with the first pulse and three interactions with the second pulse also cancel. However, those diagrams do not contain new spectral information at short population times and are therefore not of interest for the question of under what conditions new spectral information is available.

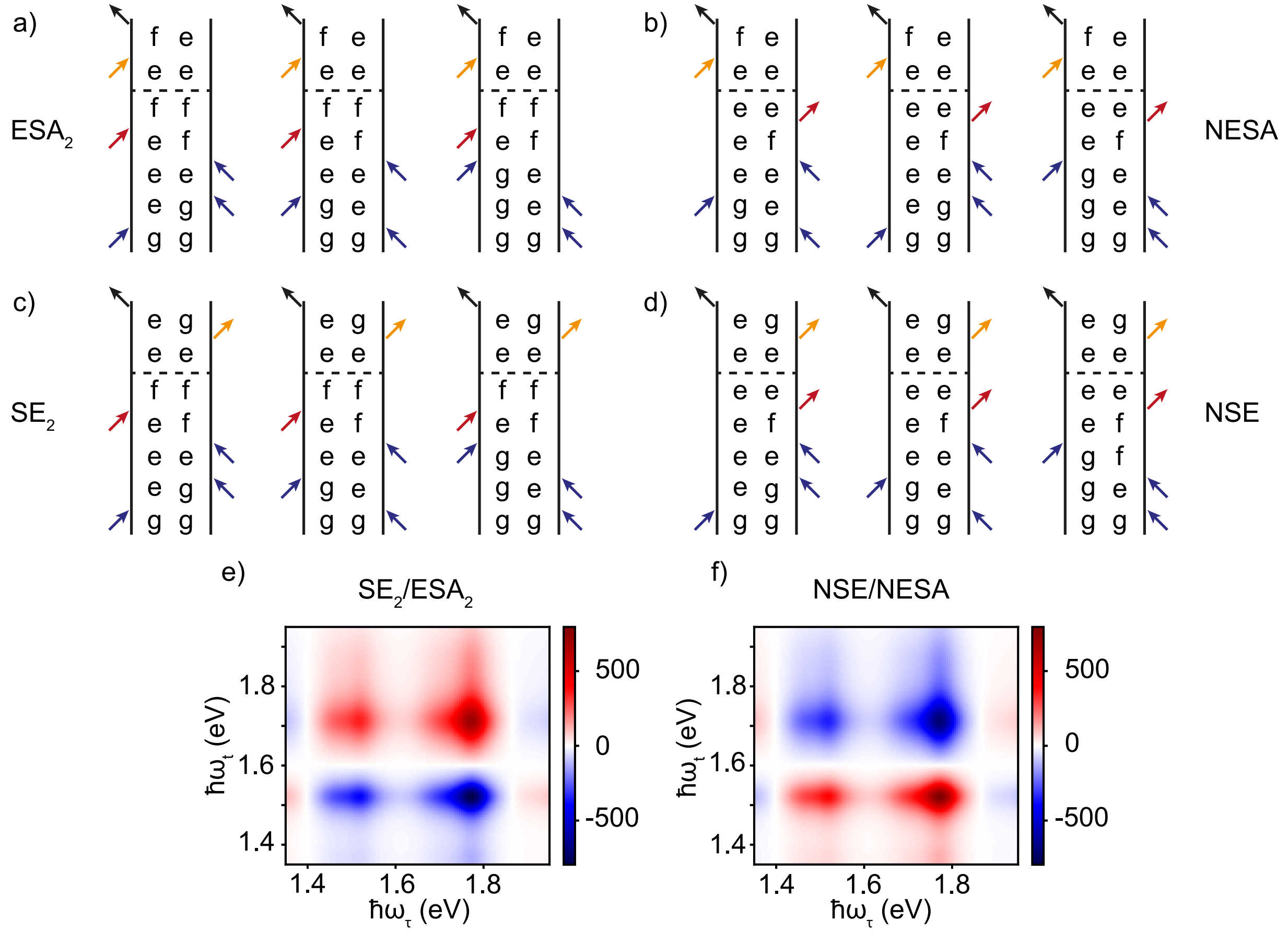


Figure S9: a) Rephasing Feynman diagrams for the $ESA_2$ process canceling b) the NESA diagrams. c) Rephasing Feynman diagrams for the $SE_2$ process canceling d) the NSE diagrams. In the diagrams, blue arrows indicate an interaction with the first pump pulse, red ones with the second pump pulse and orange ones with the probe pulse. e) Simulated 2D signals for the canceling $ESA_2$ and $SE_2$ diagrams. f) Simulated 2D signals for the canceling NESA and NSE diagrams.

## S8 Concentration-dependent linear absorption spectra

The squaraine dimer dSQBC shows small changes in relative peak intensities in the linear absorption spectrum, measured with a Jasco V-670 Spectrophotometer, for different solution concentrations, as shown in Figure S10. For the lowest-concentrated solution (blue curve in Figure S10) the peaks at $\omega_t = 1.68\,\text{eV}$ and $1.78\,\text{eV}$ show a slightly lower relative peak height compared to the main peak at $\omega_t = 1.5\,\text{eV}$. Other squaraine dyes described in the literature [6] show concentration-dependent stacking and dimerization effects. For the molecules described in [6] an increase in concentration leads to a decrease in intensity for peaks attributed to monomer and dimer species, while simultaneously new peaks arise, which correspond to aggregated molecules. For dSQBC, a further increase in concentration does not lead to further changes in relative peak intensities, as seen when comparing the linear absorption spectra for the two highest sample concentrations in Figure S10. As no new peaks are visible even for the highest concentration shown here, we assume that no significant aggregation effects are present in our sample.

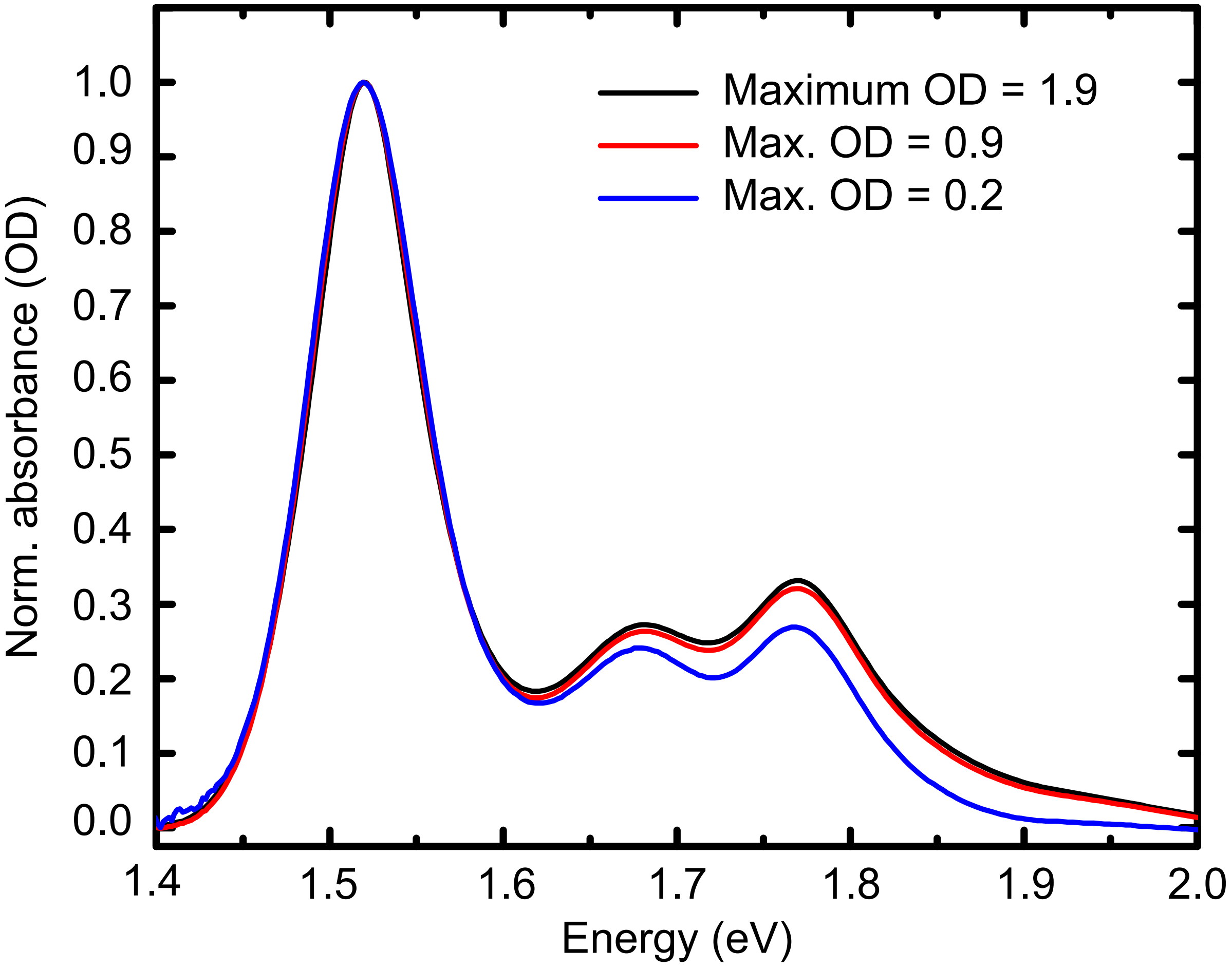


Figure S10: Normalized linear absorption spectra for differently concentrated mixtures of dSQBC and toluene. The blue curve shows the measured linear absorption spectrum for a solution with a maximum absorbance of 0.2 OD, while the red and black curves correspond to higher concentrated samples with maximum absorbance values of 0.9 OD and 1.9 OD, respectively.

## S9 Detection-energy dependence of the extraction error

To verify how the choice of the detection frequency for the intensity optimization influences the results, we analyzed the predicted total errors across all detection frequencies in two ways. To keep the analysis consistent with our experimental work we consider sets of four excitation intensities. First, we calculate the total extraction errors across the full range of detection frequencies using the fixed excitation intensities determined from the saturation parameters at $\hbar\omega_t = 1.51\,\mathrm{eV}$, which were used in the experiment. We next use the saturation parameters obtained for each detection energy to independently calculate the optimal excitation intensities and total extraction error at each detection frequency. The predicted errors for the optimal intensities at $\hbar\omega_t = 1.51\,\mathrm{eV}$ are depicted by the blue line in Figure S11, while the orange data points show the error values for the individually optimized intensities. To determine if limiting the maximal excitation intensity influences the error values, we simulated the total errors for a set with the maximal excitation intensity $I_{\max} = 15\,\mathrm{nJ}$, like in the optimization procedure for the experiments (Figure S11a) and without a maximal excitation intensity (Figure S11b).

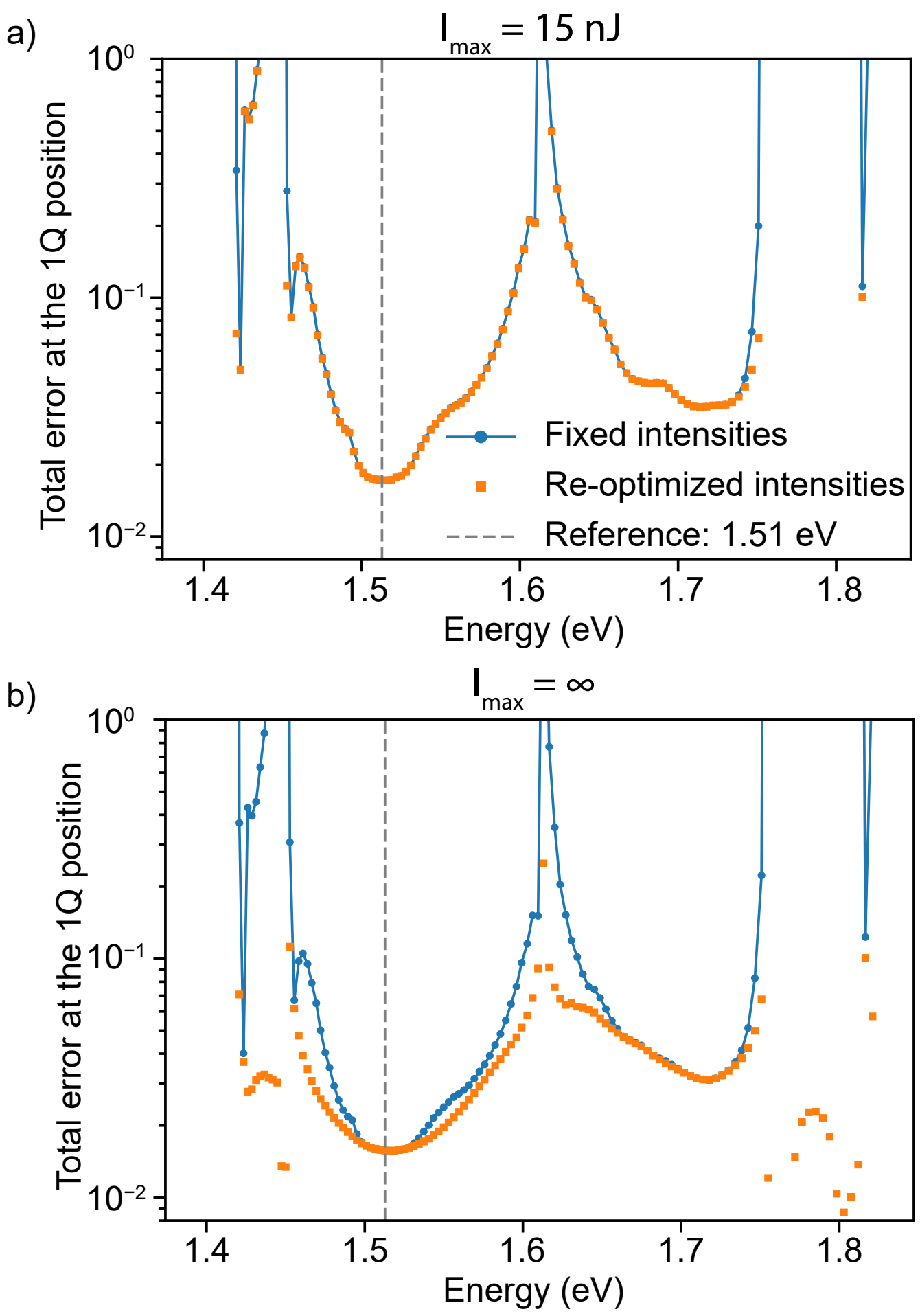


Figure S11: Total error for the order extraction at the 1Q position. The blue curve shows the total errors as a function of the detection energy for the set of intensity values optimized using the saturation parameters at a reference value $\hbar\omega_t = 1.51\,\mathrm{eV}$. The orange data points show the total extraction error obtained when the excitation intensities are optimized independently for each detection energy using the saturation parameters determined at that same energy. a) Total errors for a set maximal limit of the optimized intensities of $I_{\max} = 15\,\mathrm{nJ}$. b) Total errors without a maximal excitation intensity.

With our experimental limit $I_{\mathrm{max}} = 15\,\mathrm{nJ}$, the total errors for fixed intensities and re-optimized intensities are nearly identical (Figure S11a). Therefore, choosing a different reference energy does not influence the extraction method for our experimental setup and the investigated sample. If our experiment were not limited to a maximal excitation intensity (Figure S11b), the optimization of the errors still shows that the predicted errors in the relevant spectral regions – between $\hbar\omega_t = 1.49\,\mathrm{eV}$ and $1.6\,\mathrm{eV}$ and between $\hbar\omega_t = 1.62\,\mathrm{eV}$ and $1.73\,\mathrm{eV}$ – are similar when using the fixed intensities chosen with the reference energy of $\hbar\omega_t = 1.51\,\mathrm{eV}$ and when using individually re-optimized intensities. We can also identify three spectral regions, below $\hbar\omega_t = 1.49\,\mathrm{eV}$, around $\hbar\omega_t = 1.61\,\mathrm{eV}$, and above $\hbar\omega_t = 1.73\,\mathrm{eV}$, where reoptimizing the experimental excitation intensities would allow a significant decrease in predicted extraction errors. However, the predicted errors in those regions are not reliable; for the regions around $\hbar\omega_t = 1.61\,\mathrm{eV}$ and above $\hbar\omega_t = 1.75\,\mathrm{eV}$, the saturation fit to the TA data does not show good results, as explained in Section S3. This phenomenon occurs where the $S^{(3)}$ signal shifts from bleach to ESA features[5]. These are precisely the regions where the signal itself is small, as we can see in the 2D measurement. Thus, the differences in error shown in Figure S11 for these regions have little influence on the quality of our order extraction procedure.

Figure S11 shows that the total error up to $S^{(7)}$ (which is dominated by the error in $S^{(7)}$ itself) is estimated to be below 5% across the spectral regions we use for this analysis, except where the signal itself is small, near 1.61 eV. Small improvements to the error could be obtained, in principle, by changing intensities for each detection frequency with higher maximum pulse energies than we can achieve with out setup, but our choice to optimize at the linear absorption maximum produces reliable extractions across our spectral window.

## S10 Effect of chirped pulses on simulation results

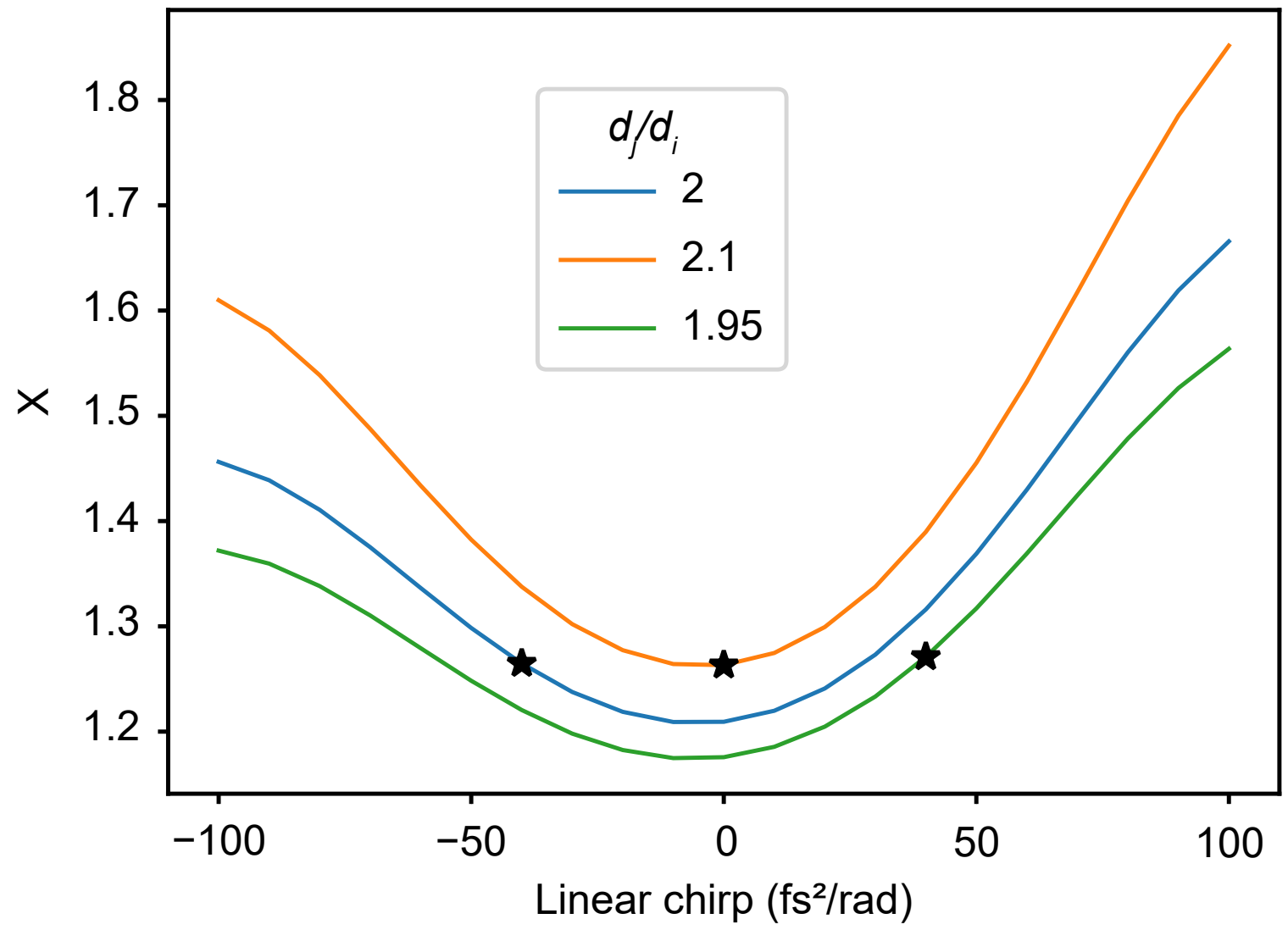


Figure S12: $X$ ratio as a function of pump pulse GDD for three different values of $d_j/d_i$. Stars indicate the optimized value of $d_j/d_i$ that gives $X = 1.25$ for positive GDD (green), negative GDD (blue) and transform-limited pulses (orange).

As described in the main manuscript, the chirp of the pump pulses can have a significant effect on the X ratio. Figure S12 shows how the X ratio varies as a function of group-delay dispersion (GDD) for fixed model parameters. We fit the phase information retrieved via FROG to determine that the pump pulse contains a GDD of $\pm 36\,\mathrm{fs}^2$. We did not obtain the sign of GDD. However, as can be seen in Figure S12, the $X$ ratio is nearly symmetric close to GDD $= 0$, and our extracted value of $d_j/d_i$ does not depend strongly on the sign. The measured electric field is shown in Figure S13, along with the fits we performed to the

intensity profile and the phase. We used fits, rather than the raw data, so that we could run convergence checks with respect to the number of time/frequency points and spacing used to resolve the electric field[1]. This convergence analysis allowed us to ensure that we did not have numerical errors, and it also allowed us to find the minimum number of points needed to produce accurate simulations, thereby making our higher-order simulations more efficient. After determining the optimal parameters to obtain X = 1.25, we ran one additional simulation using the measured electric field, instead of the fitted values. The value of X changed by less than 3%, verifying that our fits captured all of the important information. Specifically, we took the linear laser intensity measured with a spectrometer, took the square root, and multiplied by a complex exponential phase factor with the phase determined from the FROG evaluation, which we interpolated onto the frequency grid of the independent intensity measurement. We then zero-padded the spectrum to enhance the temporal resolution and took the discrete Fourier transform. Finally, we truncated the time-domain field to lie between -500 fs and +500 fs.

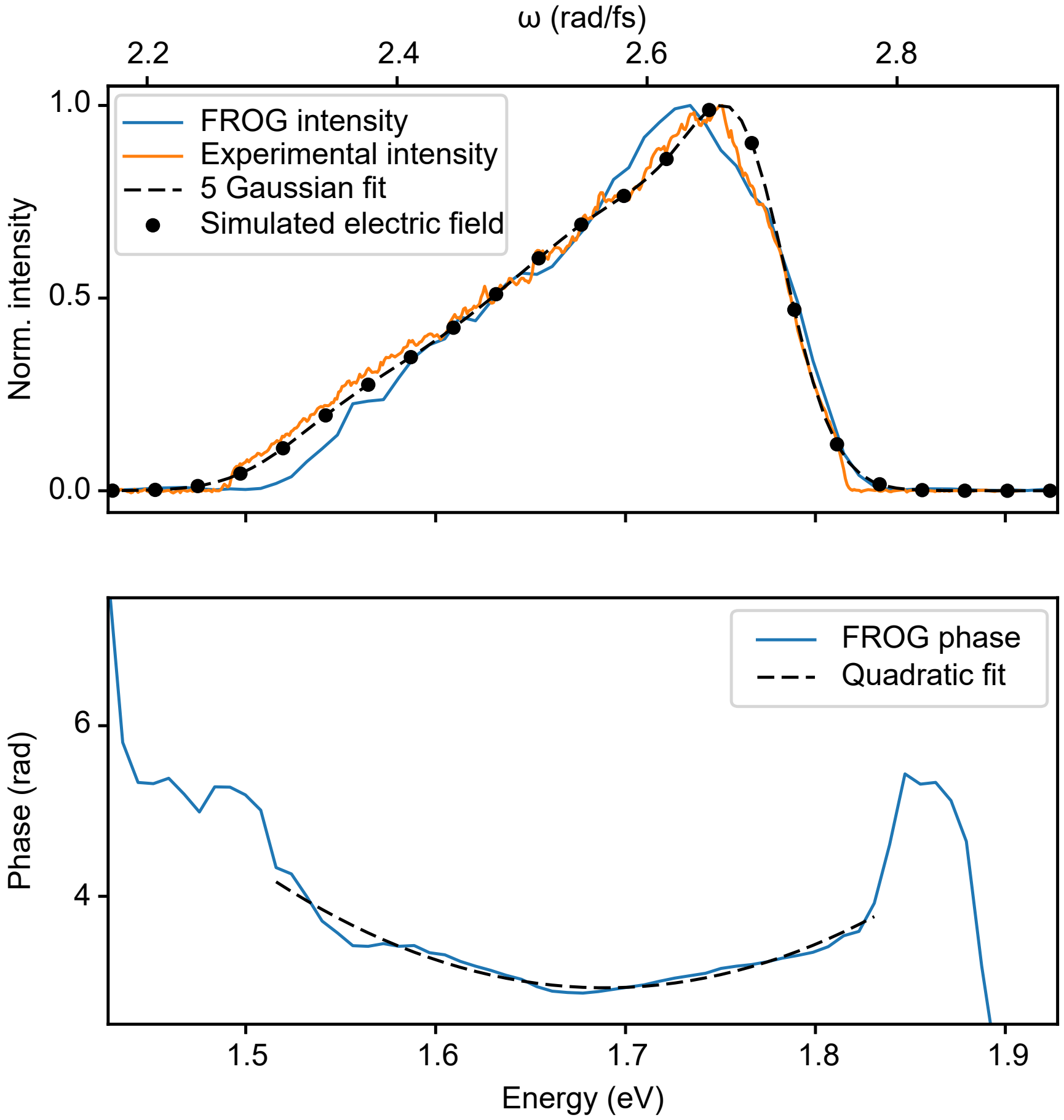


Figure S13: Measured (solid lines) and fitted (dashed lines) electric field intensity (top panel) and phase (bottom panel). The blue curve shows the intensity profile reconstructed from FROG data and the orange line shows the linear spectral intensity measured independently with a spectrometer. The grey dashed curve shows the sum of five Gaussians used as the intensity spectrum for our simulations. The black dots in the top panel represent the data points used to simulate the electric field. We fit the phase in the spectral region where the intensity was greater than 1% of the maximum.

## S11 Effect of $|j\rangle$ exciton relaxation dynamics on the $X$ ratio

In the limit where all multi-excitations have relaxed into the single-excitation manifold, and all single-excitons have relaxed to the lowest-energy exciton, $|i\rangle$, the ratios $X_{\mathrm{SE}}$ and $X_{\mathrm{ESA}}$ should be equal. Since the experimental data were collected at a population time of 100 fs, we do not expect to have reached this limit completely, and indeed $S_{\mathrm{SE}} \neq X_{\mathrm{ESA}}$. Furthermore, Figure S7 shows evidence that the third-order signal shows signs of having some remaining $|j\rangle$ exciton population. The simulations and experimental $\hbar\omega_\tau$ traces in Figure S7 match qualitatively at a simulation time of $T = 150\,\mathrm{fs}$. We see additionally in Figure S14 that both $X_{\mathrm{SE}}$ and $X_{\mathrm{ESA}}$ match the experimental values around $T = 150\,\mathrm{fs}$, but only for negative chirp. In our model,the lifetime of the $|j\rangle$ exciton is around 150 fs. This lifetime does not depend upon$\mu_{cf}$ and $\mu_{bf}$, which are the only quantities that are changed for positive and negative chirp. The underlying dynamics are still slightly different, since the excitons are populated at different times, given the different arrival times of the red and blue frequencies.

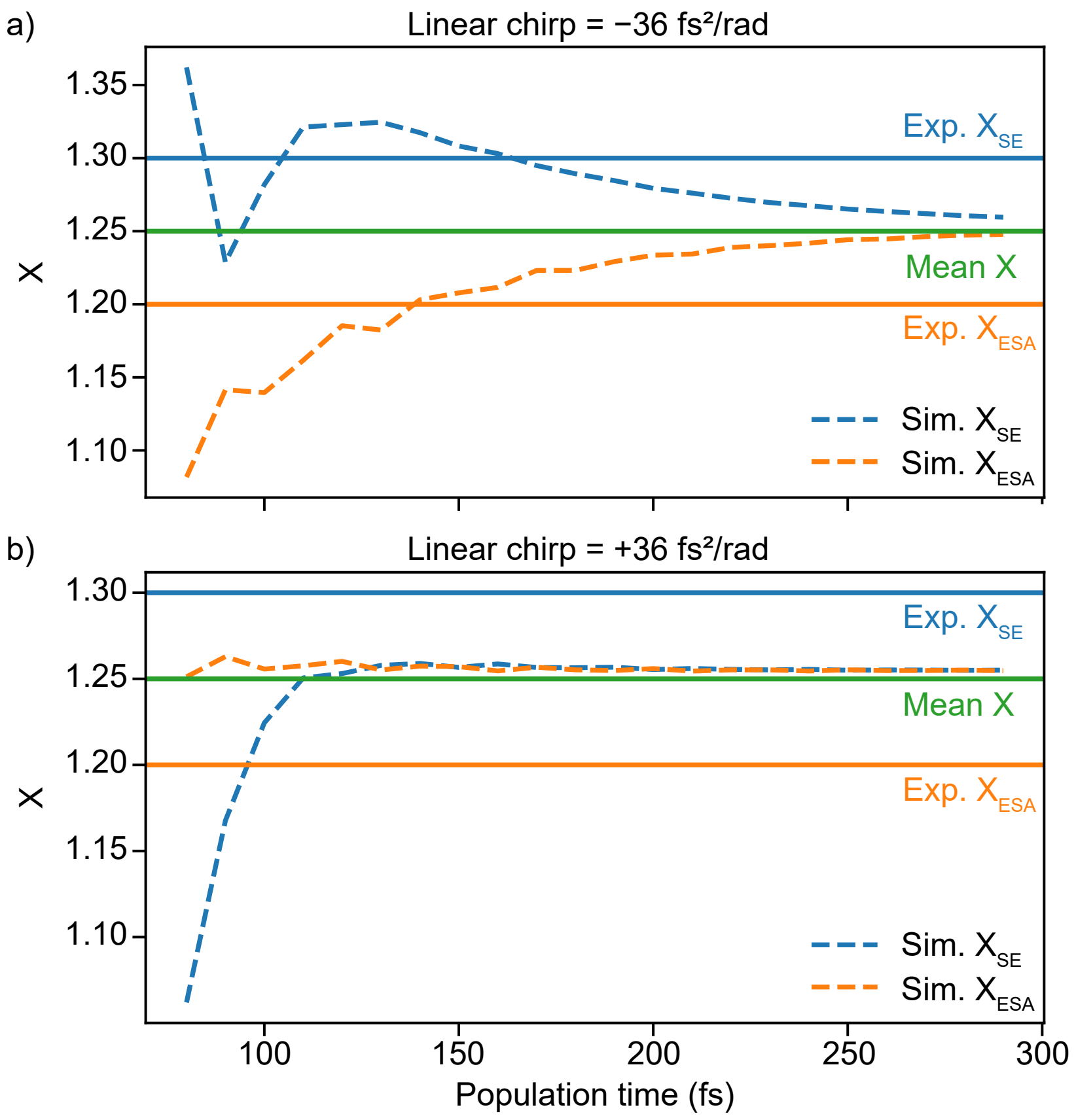


Figure S14: $X$ ratio as a function of population time for (a) negative linear chirp and (b) for positive linear chirp. Simulations are performed for the optimized parameters for the respective chirp. Although both negative and positive linear chirp relax to the same value of $X$ for long population time, they have much different population dynamics. In the case of negative linear chirp, the experimental and simulated values of $X_{\mathrm{SE}}$ and $X_{\mathrm{ESA}}$ agree around 150 fs, which corresponds to the decay lifetime of the $|j\rangle$ exciton, which decays into the $|i\rangle$ exciton.

We provide these simulations to demonstrate that remaining $|j\rangle$ excitons can cause the ratios $X_{\mathrm{ESA}}$ and $X_{\mathrm{SE}}$ to be different from one another, and also that in such a case they ultimately reach the average value that we fit to. We cannot say for certain that this is the cause of difference in the two ratios in our data, because we do not know the sign of the chirp, and also because we did not acquire many population times. In addition, it is interesting and useful to note that although the long-time limit was not strongly affected

by the sign of the chirp (see Table S2), that the transient dynamics can be strongly affected by the sign of the chirp. This makes it clear that in general, the analysis of higher-order signals requires not only the magnitude of the chirp, but also its sign.


\fi

\end{document}